\documentclass[]{iopart}

\usepackage{graphicx,amssymb}
\usepackage[usenames,dvipsnames]{xcolor}
\usepackage{lscape}
\expandafter\let\csname equation*\endcsname\relax
\expandafter\let\csname endequation*\endcsname\relax
\usepackage{amsmath}
\usepackage[retainorgcmds]{IEEEtrantools}
\usepackage{cite} 
\usepackage[symbol*]{footmisc}
\usepackage{url} 
\usepackage{hyperref}

\begin{document}

\newcounter{count}
\title[The numerical relativity breakthrough for binary black holes]{The numerical relativity breakthrough for binary black holes}

\author{Ulrich Sperhake$^{1,2,3}$
       }

\address{${}^1$ Department of Applied Mathematics and Theoretical Physics,
         Centre for Mathematical Sciences, University of Cambridge,
         Cambridge CB3 0WA, United Kingdom}

\address{${}^2$ Theoretical Astrophysics 350-17,
         California Institute of Technology, Pasadena, CA 91125}

\address{${}^3$ Department of Physics and Astronomy,
         The University of Mississippi, University, MS 38677, USA}

\ead{U.Sperhake@damtp.cam.ac.uk}

\begin{abstract}
  The evolution of black-hole binaries in vacuum spacetimes constitutes
  the two-body problem in general relativity. The solution of this
  problem in the framework of the Einstein field equations is
  a substantially more complex exercise than that of the dynamics of two
  point masses in Newtonian gravity, but it also presents us with
  a wealth of new exciting physics. Numerical methods are likely the
  only method to compute the dynamics of black-hole systems in the fully
  non-linear regime and have been pursued since the 1960s, culminating
  in dramatic breakthroughs in 2005. Here we review the methodology
  and the developments that finally gave us a solution of this
  fundamental problem of Einstein's theory and discuss the breakthrough's
  implication for the wide range of contemporary black-hole physics.
\end{abstract}


\section{Introduction}
\label{sec:intro}

The interaction of two point masses is probably the simplest
and most fundamental dynamical problem one can conceive of in
a theory of gravity. This problem is sufficiently simple in
Newton's theory such that it can be solved analytically. In spite
of the simplifications, the Newtonian two-body problem describes
with high accuracy a wide class of physical systems, ranging from
the planetary orbits in the solar system to the motion of the spacecraft
that carried humans to the moon in 1969. Observed deviations
in the motion of Uranus from the Newtonian predictions led to
the prediction of a further planet, Neptune, that was indeed identified in
1846. For a while, a similar explanation was considered for anomalies
observed in the perihelion precession of Mercury. The conjectured
planet ``Vulcan'', however, has never been found and in this case
the explanation came in the form of a {\em modified theory of gravity},
namely Einstein's general relativity (GR).

GR differs from Newtonian gravity not only in terms of quantitative
predictions but also presents a conceptually totally different
description of gravity. Acceleration of objects due to gravitational
interaction is no longer the result of a {\em force} but due to
the curvature of spacetime itself. Mathematically, the spacetime
is described in terms of a
manifold $\mathcal{M}$ equipped with a metric $g_{\alpha \beta}$
which is determined through Einstein's field equations
\begin{equation}
  R_{\alpha \beta} -\frac{1}{2}g_{\alpha \beta} R
  + \Lambda g_{\alpha \beta} = \frac{8\pi G}{c^4} T_{\alpha \beta}\,.
  \label{eq:Einstein4D}
\end{equation}
Here, Greek indices range from $0$ to $3$,
$R_{\alpha \beta}$ is the Ricci tensor, $R$ the Ricci scalar,
$\Lambda$ the cosmological constant
and $T_{\alpha \beta}$ the energy momentum tensor. Unless specified
otherwise, we will work in units where the gravitational constant and
speed of light are unity, $G=1=c$. Much of this work will focus on the
case of vacuum and asymptotically flat spacetimes where $\Lambda=0$
and $T_{\alpha \beta}=0$ and the Einstein equations become
$R_{\alpha \beta}=0$.

The Ricci tensor $R_{\alpha \beta}$ is a non-linear function of the
spacetime metric components $g_{\alpha \beta}$ and their first and
second derivatives. The Einstein equations
couple space and time in a complex manner to the gravitational sources
which is encapsulated in Wheeler's popular phrase ``Matter tells spacetime
how to curve, spacetime tells matter how to move''. As we shall
discuss in more detail in Sec.~\ref{sec:Einsteinian}, this non-linear coupling
of geometry and sources\footnote{In particular the fact
that gravity itself represents
energy and, thus, a source of gravity, accounts for the richness of
vacuum solutions in GR.} makes the two-body problem much more
complicated in GR but also lends a vast richness of new physics to
these seemingly simple systems. Most importantly, we have the following
conceptual differences between the Newtonian and the general relativistic
case. (i) Sources of finite
mass-energy cannot be point-like in GR but inevitably
represent extended regions of non-vanishing curvature. The closest
approximation to a point-like source in GR is a black hole (BH)
and the two-body problem in GR therefore is a BH binary. (ii) The interaction
of the two BHs is dissipative as energy and momentum can be radiated away
from the binary in the form of gravitational waves (GW) which
are subject of large-scale efforts for direct detection.
Bound systems
therefore eventually result in the merger of the two constituents.

In view of these special features of GR, the BH binary problem is
often regarded as a three-stage process: (i) an extended phase of
the interaction of two separate BHs, often referred to
as the {\em inspiral} for bound systems, (ii) the
{\em merger}, and (iii) the {\em ringdown}, a process of damped
sinusoidal oscillations as the post-merger
remnant sheds all structure beyond mass and angular momentum
and settles down into a stationary Kerr BH. Unbound systems do
not undergo stage (ii) and (iii) of this process but may still
interact in a highly non-linear manner during stage (i).
The challenge to accurately model all possible stages of the dynamics
of a BH binary then consists in solving Einstein's vacuum equations
$R_{\alpha \beta}=0$, a system of 10 coupled, non-linear second-order
partial differential equations (PDEs). This challenge has often been referred
to as
the {\em Holy Grail} of numerical relativity (NR) and how it has eventually
been met is the subject of this review.

For understanding the magnitude of this challenge and the particular
issues arising in GR, it will
be instructive to contrast it with its Newtonian counterpart and
we will therefore start in Sec.~\ref{sec:Newtonian} with a brief review of the
Newtonian two-body problem. The GR case is then summarized in
Sec.~\ref{sec:Einsteinian} including a ``todo list'' of items that specifically
arise in solving Einstein's rather than Newton's equations.
The methodology to address these items is discussed in Sec.~\ref{sec:NR}.
We continue in Sec.~\ref{sec:History}
with an overview of the historical progress of the
community which culminated in the 2005 breakthroughs by
Pretorius \cite{Pretorius:2005gq} as well as
the {\em moving puncture} method of the
Brownsville (now Rochester) \cite{Campanelli:2005dd}
and the NASA Goddard \cite{Baker:2005vv} groups who,
quite remarkably, presented two rather different methods to
solve the BH binary problem within some months.
In Sec.~\ref{sec:Morphology} we summarize the physical
features of the dynamics of BH binary systems and briefly
discuss the importance of the GR two-body problem in contemporary
physics. We conclude in Sec.~\ref{sec:conclusions} where we also
provide references for further reading.


\section{The Newtonian two-body problem}
\label{sec:Newtonian}

\begin{figure}[t]
  \centering
  \includegraphics[height=100pt,clip=true,angle=-180]{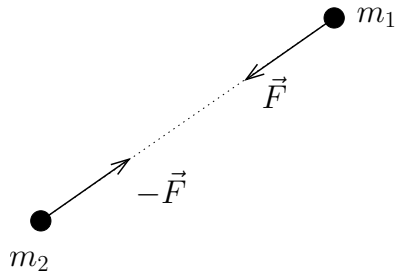}
  \caption{Illustration of the Newtonian two-body problem.}
  \label{fig:Newtonian}
\end{figure}

In the Newtonian two-body problem, we consider two point masses $m_1$
and $m_2$ moving in a background space and time.
The two masses are
separated by a distance vector $\vec{r} \equiv \vec{r}_1-\vec{r}_2$, and
we denote by $\vec{F}$ the gravitational force exerted by $m_2$ on
$m_1$. By Newton's laws $m_1$ acts on $m_2$ with $-\vec{F}$
and we have
\begin{equation}
  \vec{F} = - \frac{Gm_1 m_2}{r^2} \hat{\vec{r}}\,,
\end{equation}
where $r\equiv |\vec{r}|$ and $\hat{\vec{r}} \equiv \vec{r}/r$ is the
unit vector pointing from $m_2$ to $m_1$; cf.~Fig.~\ref{fig:Newtonian}.
The equations of motion for
the two particles are given by
\begin{equation}
  m_1 \frac{d^2 \vec{r}_1}{dt^2} = \vec{F} = -G \frac{m_1 m_2}{r^2}
        \hat{\vec{r}} = - m_2 \frac{d^2 \vec{r_2}}{dt^2}\,.
  \label{eq:Nmotion}
\end{equation}
These equations are most conveniently solved by introducing the
{\em reduced mass} $\mu = m_1 m_2 / (m_1+m_2)$ and rewriting
(\ref{eq:Nmotion}) as the equation of motion for a single particle
of mass $\mu$,
\begin{equation}
  \mu \frac{d^2 \vec{r}}{dt^2} = \vec{F}\,.
\end{equation}

Without loss of generality, we choose Cartesian coordinates $x,\,y,\,z$
such that the particles' motion takes place in the plane $z=0$
and we furthermore introduce polar coordinates $r,\,\theta$ with
$x=r\cos \theta$, $y=r\sin \theta$. We thus obtain two constants
of motion, the energy $E$ and the angular momentum $L$ given by
\begin{eqnarray}
  E &=& \frac{1}{2}\mu \left[ \left( \frac{dr}{dt} \right)^2
        + r^2 \left( \frac{d\theta}{dt} \right)^2 \right]
        - G\frac{m_1 m_2}{r}\,,
  \label{eq:NE} \\[10pt]
  L &=& \mu r^2 \frac{d\theta}{dt}\,.
  \label{eq:NL}
\end{eqnarray}
By substituting $d\theta/dt$ in (\ref{eq:NE}) in terms of $L$ through
(\ref{eq:NL}) and solving the two equations for $dr/dt$ and
$d\theta/dt$, respectively, we obtain a differential equation for
$r$ regarded now as a function of $\theta$:
\begin{equation}
  \frac{dr}{d\theta} = \frac{\dot{r}}{\dot{\theta}}
        = \frac{\sqrt{2\mu} r^2}{L} \sqrt{E - \frac{L^2}{2\mu r^2}
          + G \frac{m_1 m_2}{r}}\,,
  \label{eq:Nfinal}
\end{equation}
where a ``dot'' denotes a time derivative $d/dt$. A solution to
Eq.~(\ref{eq:Nfinal}) is given in closed analytic form by
\begin{equation}
  r = \frac{r_0}{1+\epsilon \cos \theta}\,,~~~~
  r_0 = \frac{L^2}{\mu G m_1 m_2}\,,~~~~
  \epsilon = \sqrt{1+ \frac{2EL^2}{\mu (Gm_1 m_2)^2}}\,,
  \label{eq:Nsol}
\end{equation}
where the {\em semilatus rectum} $r_0$ and the {\em eccentricity} $\epsilon$
are determined completely in terms of the constants of motion $E$, $L$.

The solutions to Eq.~(\ref{eq:Nfinal}) can be classified into the
following four types.
\begin{list}{\rm{(\arabic{count})}}{\usecounter{count}
             \labelwidth1cm \leftmargin1.5cm \labelsep0.4cm \rightmargin1cm
             \parsep0.5ex plus0.2ex minus0.1ex \itemsep0ex plus0.2ex}
  \item {\bf \em Circular orbits} given by
        $\epsilon =0$ where $E$ takes on its minimal possible value
        $E_{\rm min}$
        and the solutions are circles $r(\theta) = r_0$.
  \item {\bf \em Kepler ellipses} given by $0 < \epsilon < 1$
        or, equivalently $E_{\min} < E < 0$. In this case, the
        solution (\ref{eq:Nsol}) can be written as
        \begin{equation}
          \frac{(1-\epsilon^2)^2}{r_0^2} \left( x + \frac{\epsilon r_0}
                {1-\epsilon^2} \right)^2
                + y^2 \frac{1-\epsilon^2}{r_0^2} = 1\,,
        \end{equation}
        which is of the general form $(x-x_0)^2/a^2 + (y-y_0)^2/b^2=1$
        for an ellipse centered on $(x_0,y_0)$.
  \item {\bf \em Parabola} given by $\epsilon = 1~~\Leftrightarrow~~E=0$
        in which case (\ref{eq:Nsol}) takes on the form
        $2r_0 x+ y^2 = r_0^2$.
  \item {\bf \em Hyperbolic orbits} given by $\epsilon>1~~\Leftrightarrow~~
        E>0$ where the solution (\ref{eq:Nsol}) can be written as
        $-(\epsilon^2-1) x^2 + 2r_0 \epsilon x + y^2= r_0^2$.
\end{list}
The elliptic type of solutions is often extended to include the circular
case (1) and we have the three classic cone cross sections of
elliptic, parabolic and hyperbolic particle curves.

\section{The general relativistic two-body problem}
\label{sec:Einsteinian}

We have seen that the Newtonian two-body problem
can be formulated as one ordinary differential equation
(\ref{eq:Nfinal}) for which initial data at $t=0$
need to be specified in the
form of the initial position $(r,\theta)$ and
the velocity components $\dot{r}$ and $\dot{\theta}$ or,
alternatively, as $dr / d\theta$. The free parameters of the system
are given by the masses $m_1$ and $m_2$ as well as the
constants of motion $E$ and $L$.

In order to illustrate the many fundamental differences that arise in
the general relativistic two-body problem, it is helpful to first 
consider the Einstein equations (\ref{eq:Einstein4D}) in a time-space
split form. This is conveniently achieved with
the canonical ``3+1'' split of the Einstein equations
originally developed by Arnowitt, Deser and Misner (ADM) \cite{Arnowitt:1962hi}
and later reformulated by York \cite{York1979,York1982};
for a detailed review see \cite{Gourgoulhon:2007ue}. We consider for
this purpose a manifold $\mathcal{M}$ equipped with a spacetime metric
$g_{\alpha \beta}$ and assume that there exists a
function $t :\mathcal{M} \rightarrow \mathbb{R}$
that satisfies the following two properties. (i) The 1-form
$\boldsymbol{\mathsf{d}}t$
is timelike everywhere, and (ii) the hypersurfaces
$\Sigma_t$ defined by $t = \mathrm{const}$ are non-intersecting
and $\cup_{t\in \mathbb{R}} \Sigma_t = \mathcal{M}$. The resulting
sequence of hypersurfaces $\Sigma_t$ is often referred to as a {\em foliation}
of the spacetime and we denote by $\boldsymbol{n} \equiv -
\boldsymbol{\mathsf{d}}t / ||\boldsymbol{\mathsf{d}}t||$ the
future pointing unit normal field
\begin{figure}
  \centering
  \includegraphics[height=300pt,clip=true,angle=-90]{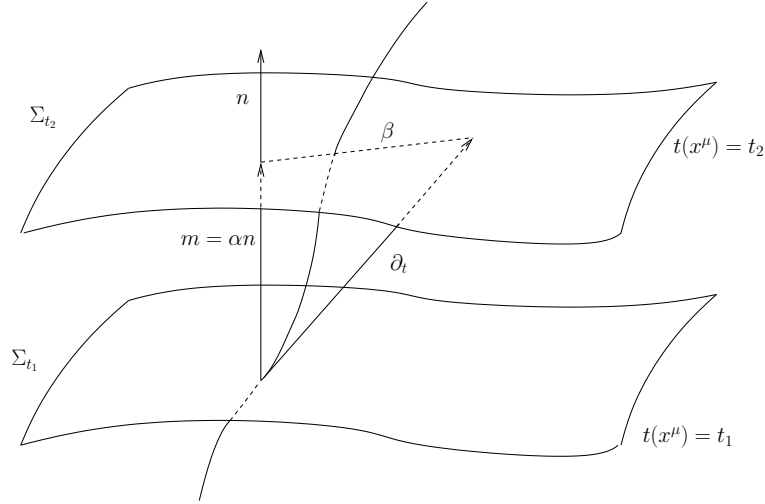}
  \caption{Two hypersurfaces of a foliation $\Sigma_t$ are shown.
           The lapse $\alpha$ and shift vector $\boldsymbol{\beta}$
           relate the unit normal field $\boldsymbol{n}$ to the coordinate
           vector $\boldsymbol{\partial}_t$. Note that
           $\langle \boldsymbol{\mathsf{d}}t,\alpha \boldsymbol{n}
           \rangle = 1$ and, hence, the shift vector is
           tangent to $\Sigma_t$.
          }
  \label{fig:foliation}
\end{figure}
of the $\Sigma_t$. We furthermore define coordinates $x^{\alpha}$
to be {\em adapted} to the foliation if $x^0 = t$ and the $x^i$,
$i=1, \ldots, 3$, form a coordinate system in each hypersurface
$\Sigma_t$.

It turns out convenient to define the {\em lapse function} and
{\em shift vector} by
\begin{equation}
  \alpha \equiv \frac{1}{|| \boldsymbol{\mathsf{d}} t ||}\,,~~~~~~~~
  \beta^{\mu} \equiv (\partial_t)^{\mu} - \alpha n^{\mu}\,.
\end{equation}
Lapse and shift relate the unit normal direction $\boldsymbol{n}$
to the direction $\boldsymbol{\partial}_t$ of the coordinate time $t$
which is illustrated in Fig.~\ref{fig:foliation}. One straightforwardly
shows that $\langle \boldsymbol{\mathsf{d}}t, \alpha \boldsymbol{n}
\rangle = 1$ and, together with $\langle \boldsymbol{\mathsf{d}}t,
\boldsymbol{\partial}_t \rangle = 1$, it follows that the
shift vector $\boldsymbol{\beta}$ is tangent to $\Sigma_t$. Finally,
the lapse function relates proper time $\tau$ as measured by an observer
with four-velocity $n^{\alpha}$ to the coordinate time $t$:
$\Delta \tau = \alpha\, \Delta t$.

Having decomposed the spacetime into a one-parameter family
of spatial hypersurfaces, we next consider projections of tensors.
For this purpose, we define the projection operator
$\bot^{\alpha}{}_{\mu} \equiv \delta^{\alpha}{}_{\mu} + n^{\alpha}n_{\mu}$
and the projection of an arbitrary tensor
$T^{\mu_1 \mu_2 \ldots}{}_{\nu_1 \nu_2 \ldots}$ by
\begin{equation}
  (\bot T)^{\alpha_1 \alpha_2 \ldots}{}_{\beta_1 \beta_2 \ldots}
        \equiv
        \bot^{\alpha_1}{}_{\mu_1} \bot^{\alpha_2}{}_{\mu_2} \ldots
        \bot^{\nu_1}{}_{\beta_1} \bot^{\nu_2}{}_{\beta_2} \ldots
        T^{\mu_1 \mu_2 \ldots}{}_{\nu_1 \nu_2 \ldots}\,.
\end{equation}
In particular, the spatial projection of the metric gives us the
{\em first fundamental form} or {\em spatial metric}
\begin{equation}
  \gamma_{\alpha \beta} \equiv \bot^{\mu}{}_{\alpha}
        \bot^{\nu}{}_{\beta} g_{\mu \nu}
        = g_{\alpha \beta} + n_{\alpha} n_{\beta}
        = \bot_{\alpha \beta}\,.
\end{equation}
$\boldsymbol{\gamma}$ and $\boldsymbol{\bot}$ thus represent the same
tensor and we shall use both symbols depending on whether the emphasis
is on the projection operation or the geometry of the spatial slices.
It is straightforward to show that the components of the spacetime
metric in adapted coordinates are related to the spatial metric, lapse
and shift according to
\begin{equation}
  g_{\alpha \beta} = \left(
        \begin{array}{c|c}
          -\alpha^2 + \beta_m \beta^m & \beta^j \\
          \hline
          \beta^i & \gamma_{ij}
        \end{array}
        \right)
        ~~~\Leftrightarrow~~~
  g^{\alpha \beta} = \left(
        \begin{array}{c|c}
          -\alpha^{-2} & \alpha^{-2} \beta^j \\
          \hline
          \alpha^{-2} \beta^i & \gamma^{ij}
          - \alpha^{-2} \beta^i \beta^j
        \end{array}
        \right)\,.
  \label{eq:3+1metric}
\end{equation}
Here, Latin indices $i,\,j,\,\ldots$ extend from $1$ to $3$ and spatial
indices are raised and lowered with the spatial metric $\gamma_{ij}$
and its inverse $\gamma^{ij}$. The spatial metric furthermore defines
a unique torsion-free and metric-compatible connection
$\Gamma^{i}_{jk} = \frac{1}{2} \gamma^{im}(\partial_j \gamma_{km}
+ \partial_k \gamma_{mj} - \partial_m \gamma_{jk})$ and an associated
covariant derivative for arbitrary spatial tensors given by
\begin{equation}
  D_{\gamma} S^{\alpha_1 \alpha_2 \ldots}{}_{\beta_1 \beta_2 \ldots}
        = \bot^{\lambda}{}_{\gamma} \bot^{\alpha_1}{}_{\mu_1}
        \bot^{\alpha_2}{}_{\mu_2} \ldots \bot^{\nu_1}{}_{\beta_1}
        \bot^{\nu_2}{}_{\beta_2} \ldots
        \nabla_{\lambda} S^{\mu_1 \mu_2 \ldots}{}_{\nu_1 \nu_2 \ldots}\,.
\end{equation}
The final ingredient we shall need in the space-time split of the
Einstein equations is the {\em second fundamental form} or
{\em extrinsic curvature}
\begin{equation}
  K_{\alpha \beta} = -\bot \nabla_{\beta} n_{\alpha}\,.
  \label{eq:defK}
\end{equation}
Here, the minus sign is a common convention in NR but
the extrinsic curvature is sometimes also defined with a plus sign
in the literature. Furthermore, the definition (\ref{eq:defK})
implies the relation $K_{\alpha \beta} = -\frac{1}{2}
\mathcal{L}_{\boldsymbol{n}} \gamma_{\alpha \beta}$, where $\mathcal{L}$
denotes the Lie derivative. It turns out convenient to also introduce
the following projections of the energy momentum tensor
\begin{eqnarray}
  &\rho = T_{\mu \nu}n^{\mu} n^{\nu}\,,~~~~~~~~~~~~~~~
  &j_{\alpha} = -\bot^{\nu}{}_{\alpha} T_{\mu \nu}n^{\mu}, \\
  & S_{\alpha \beta} = \bot^{\mu}{}_{\alpha} \bot^{\nu}{}_{\beta} T_{\mu \nu}\,,
  & S = \gamma^{\mu \nu} S_{\mu \nu}\,.
\end{eqnarray}
The space-time split of the Einstein equations $G_{\alpha \beta}
= 8\pi T_{\alpha \beta}$ is then obtained
through a lengthy calculation whose details can be found for example
in \cite{Gourgoulhon:2007ue}. This calculation gives six second-order
in time evolution equations as well as the Hamiltonian and three momentum
constraints
\begin{eqnarray}
  \partial_t \gamma_{ij} &=& \beta^m \partial_m \gamma_{ij}
        + \gamma_{mj} \partial_i \beta^m + \gamma_{im}\partial_j \beta^m
        - 2\alpha K_{ij}\,,
        \label{eq:dtgamma} \\
  \partial_t K_{ij} &=& \beta^m \partial_m K_{ij}
        + K_{mj} \partial_i \beta^m + K_{im} \partial_j \beta^m
        - D_i D_j \alpha
        \nonumber \\
     && + \alpha (\mathcal{R}_{ij} + KK_{ij}
        - 2K_{im} K^m{}_j)
        + 4\pi \alpha [(S-\rho) \gamma_{ij} - 2S_{ij}]\,,
        \label{eq:dtK} \\
        0 &=& \mathcal{R} + K^2 - K^{mn}K_{mn}- 16\pi \rho\,,
        \label{eq:Ham}
        \\
        0 &=& D_i K - D_m K^m{}_i + 8\pi j_i\,.
        \label{eq:mom}
\end{eqnarray}
Here, $\mathcal{R}_{ij}$ and $\mathcal{R}$ denote the Ricci tensor and
scalar associated with $\gamma_{ij}$. Note that we assume here
coordinates $(t,\,x^i)$ adapted to the foliation and therefore
have replaced spacetime indices $\alpha,\,\beta,\,\ldots$ with
spatial indices $i,\,j,\,\ldots\;$. We also see that the extrinsic
curvature $K_{ij}$ allows us to write the second-order-in-time
evolution equations as a first-order system. Finally, the constraint
equations (\ref{eq:Ham}) and (\ref{eq:mom}) are preserved under the
evolution equations because of the Bianchi identities.

At this point it is worth taking a break to consider our situation
in comparison with the Newtonian two-body problem. In place of one
ordinary differential equation, we now have a system of
coupled PDEs which forms an
initial-boundary-value problem (IBVP). This system consists of
six second-order in time evolution equations written in
Eqs.~(\ref{eq:dtgamma}), (\ref{eq:dtK}) in first-order form
as well as four constraints (\ref{eq:Ham}), (\ref{eq:mom}). Even though
the constraints are preserved under the evolution equations in
the continuum limit, care needs to be taken that constraint violations
due to numerical inaccuracies do not grow out of bounds. Furthermore, the
initial data need to satisfy the constraints which requires solving
a set of elliptic differential equations. Note
that the Einstein equations in ADM form (\ref{eq:dtgamma})-(\ref{eq:mom})
make no predictions about the lapse function $\alpha$ and the
shift vector $\beta$. Instead, these functions represent the
{\em diffeomorphism invariance} or {\em gauge freedom} of general
relativity; they can be freely specified to fix
the coordinates but, as it turns out, it is highly non-trivial
to find gauge conditions that ensure numerically stable evolutions.

It is instructive to count the number of physical degrees of freedom
contained in the system (\ref{eq:dtgamma})-(\ref{eq:mom}). We have
ten components of the Einstein metric $g_{\alpha \beta}$ corresponding
to the ten functions $\gamma_{ij}$, $\beta^i$ and $\alpha$ in the
ADM formulation. Four of these, the lapse and shift, are freely specifiable
and do not contain physical information. The constraints impose four
further conditions on the remaining functions $\gamma_{ij}$ that
must be satisfied on each hypersurface $\Sigma_t$ and we are left
with two gravitational degrees of freedom which correspond to the
$+$ and $\times$ GW polarization modes; see e.g.~\cite{Sathyaprakash:2009xs}.
The two gravitational degrees of freedom are recovered even more
elegantly in the characteristic formulation of the Einstein equations
developed by Bondi, Sachs and collaborators \cite{Bondi:1962px,Sachs:1962wk}.
Here, one chooses at least one coordinate to be null and thus foliates
spacetime in terms of light cones. The Einstein equations assume a natural
hierarchy of 2 evolution equations, 4 hypersurface equations relating
variables inside the hypersurfaces, 3 supplementary and 1 trivial equation;
for details see \cite{Winicour:2005ge} and references therein. Codes based
on the characteristic formulation have been applied with great success
in the presence of special spacetime symmetries and indeed been the
first to model single BH spacetimes with long-term stability
\cite{Gomez:1998uj,Lehner:1998ti}. In spite of the formalism's appealing
properties, however, characteristic codes have as yet not been successfully
generalized to BH binaries because the formation of caustics causes
a breakdown of the coordinate system. It remains to be seen whether
this obstacle can be overcome in future investigations; for a
recent study see \cite{Babiuc:2013rra}.

In the case of a non-vanishing energy momentum tensor
$T_{\alpha \beta}$, there may be additional matter degrees of
freedom. We also note that BH spacetimes with a BH mass $M$
contain various different
length scales, the BH horizon which has a size
of\footnote{It is common practice in NR to measure
length and time in units of the BH mass $M$ which is readily converted
into SI units through the convention $c=1=G$ once a value for the mass
has been specified.} $\mathcal{O}(M)$,
the wave length of GW signals, typically of the order $\mathcal{O}(10^2~M)$,
and the wave zone of $\mathcal{O}(10^3~M)$ where perturbation theory
permits a precise definition of GWs. Finally, we need to specify outer
boundary conditions such that there is no ingoing gravitational radiation
from infinity.

Bearing in mind all these features of the Einstein equations, we face
the following list of tasks to obtain stable, accurate numerical simulations
of the binary BH problem in general relativity.
\begin{itemize}
  \item Formulate the Einstein equations in a manner that admits a
        {\em well-posed} IBVP, i.e.~ensures a continuous dependence of the
        spacetime solution on the initial data.
  \item Choose numerically suitable gauge conditions.
  \item Discretize the resulting PDEs for
        a computer based treatment.
  \item Specify physically correct boundary conditions that also satisfy
        the constraints.
  \item Find a numerical treatment of the singularities
        inherent in BH spacetimes that avoids the appearance of
        {\em non-assigned numbers}.
  \item Calculate initial data which satisfy the constraints and
        represent a realistic snapshot of the initial state
        of the physical system under consideration.
  \item Implementation of {\em mesh refinement} or similar methods
        using multiple domains to accurately handle the different
        length scales and parallelize the resulting algorithms for
        multi-processor computation.
  \item Extract physical results in a gauge-invariant manner from the
        numerical data.
\end{itemize}
In the next section we will discuss the most important methods which
have been developed for handling these tasks and have made possible the
NR breakthroughs in solving the binary BH problem in
general relativity.

\section{The ingredients of numerical relativity}
\label{sec:NR}

The techniques for addressing the above list of tasks
have been developed over several decades
through an interplay of numerical experiments and
theoretical studies carried out by numerous groups and researchers.
Many numerical investigations, especially the earlier ones, were
performed without a comprehensive understanding of all
the difficulties associated with this list of tasks. In hindsight, it is
therefore not too surprising that they met with limited success.
And yet, as is the nature of scientific exploration, all these attempts
taught us valuable lessons and contributed to the gradual assembly of
the complete picture we are going to describe in this section. We shall
present this more technical description not in chronological order
but, for reasons of clarity, topic by topic. A brief historical
review of the applications will be given in Sec.~\ref{sec:History}
below. \\[5pt]
%

\subsection{Formulations of the Einstein equations}
Any successful
attempt at numerically solving the Einstein equations must be based
on a {\em well-posed} IBVP, i.e.~a computational algorithm that
results in a time evolution which depends continuously on the initial
data. Given the inevitability of numerical noise present in
the form of round-off error in any numerical initial data, algorithms
that do not meet this criterion are evidently unsuitable for obtaining
reliable results. The well-posedness of a numerical implementation
depends on many aspects including the specific formulation of the
differential equations, gauge and boundary conditions and the
discretization schemes. Here we are concerned with the conditions
a formulation of the Einstein equations must satisfy to admit
a well-posed IBVP.

The suitability of a formulation is commonly studied in the form
of the {\em hyperbolicity} properties of the PDEs
which are related to the symmetrizeability of the {\em principal
symbol} (in simple words, the coefficient matrix of the terms
containing the highest derivatives) of the system of PDEs.
A system is called {\em strongly hyperbolic} if the principal
symbol has only imaginary eigenvalues and a complete set of linearly
independent eigenvectors \cite{Nagy:2004td}.
If the eigenvectors are not linearly
independent, the system is called {\em weakly hyperbolic}.
A system is called {\em symmetric hyperbolic} if there exists a
conserved, positive energy norm. We skip the technical details
here, but the interested reader will find extended discussions
in \cite{Reula:1998ty,Sarbach:2012pr} and references therein.
For us, the most important conclusions are the following.
(i) Of the three notions of weak, strong and
symmetric hyperbolicity, each is a stronger condition
than the previous one; cf.~Sec.~3.1.4 in \cite{Sarbach:2012pr}.
(ii) Strong hyperbolicity is a necessary
condition for a well-posed IBVP \cite{Taylor1981, Taylor1991}. (iii)
The ADM evolution equations (\ref{eq:dtgamma}), (\ref{eq:dtK})
have been shown to be weakly but not strongly hyperbolic for
fixed gauge \cite{Nagy:2004td} and a first-order version of the
ADM equations has been shown to be only weakly hyperbolic in
\cite{Kidder:2001tz}. While these studies do not constitute a rigorous proof
of the unsuitability of the ADM equations for numerical evolutions,
they strongly suggest a search for alternative formulations for which
strong hyperbolicity can be established.

Explorations of modifications of the ADM equations or alternative
formulations for use in NR already began in the late
1980s, before the full impact of the hyperbolicity properties
of the different formulations
had been realized. Over the course
of the ensuing 25 years, a great variety of different formulations
has been developed and implemented in numerical codes; for an
overview see for example \cite{Shinkai:2008yb} and in particular
Fig.~4 therein.
Here, we shall focus on those two formulations that underlie the numerical
relativity breakthroughs of 2005.

The Baumgarte-Shapiro-Shibata-Nakamura (BSSN) formulation
\cite{Nakamura:1987zz,Shibata:1995we,Baumgarte:1998te} is directly
derived from the ADM equations but works with conformally rescaled
variables, a trace split of the extrinsic curvature and promotes the
contracted Christoffel symbols to the status of independent variables.
Specifically,
the BSSN variables $\chi$, $\tilde{\gamma}_{ij}$, $K$, $\tilde{A}_{ij}$
and $\tilde{\Gamma}^i$ are defined by
%
\begin{equation}
  \chi = \gamma^{-1/3}\,,~~
  \tilde{\gamma}_{i j} = \chi \gamma_{i j}\,,~~
  K = \gamma^{m n} K_{m n}\,,~~
  \tilde{A}_{i j} = \chi \left( K_{i j} - \frac{1}{3}\gamma_{i j} K \right)\,,
  ~~
  \tilde{\Gamma}^{i} = \tilde{\gamma}^{m n} \tilde{\Gamma}^{i}_{m n}\,,
  \label{eq:BSSNvars}
\end{equation}
where $\gamma \equiv \det \gamma_{ij}$ and
$\tilde{\Gamma}^i_{jk}$ are the Christoffel symbols associated with
the conformal metric $\tilde{\gamma}_{ij}$. The BSSN system has also been
used with $\chi$ replaced by the variables $\phi = -(\ln\,\chi)/4$
or $W=\sqrt{\chi}$. Inserting the definition (\ref{eq:BSSNvars})
into the ADM equations (\ref{eq:dtgamma}), (\ref{eq:dtK}) and using
the Hamiltonian and momentum constraints respectively in the
resulting evolution equations for $K$ and $\Gamma^i$ yields
\begin{eqnarray}
  \partial_t \chi &=& \beta^m \partial_m \chi
        + \frac{2}{3} \chi (\alpha K - \partial_m \beta^m)\,,
        \label{eq:BSSNdtchi} \\
  \partial_t \tilde{\gamma}_{ij} &=& \beta^m \partial_m
        \tilde{\gamma}_{ij}
        + 2\tilde{\gamma}_{m(i} \partial_{j)} \beta^m
        - \frac{2}{3}\tilde{\gamma}_{ij} \partial_m \beta^m
        - 2\alpha \tilde{A}_{ij}\,,\\
  \partial_t K &=& \beta^m \partial_m K
        - \chi \tilde{\gamma}^{mn} D_m D_n \alpha
        + \alpha \tilde{A}^{mn}\tilde{A}_{mn}
        + \frac{1}{3}\alpha K^2 \nonumber \\
     && + 4\pi \alpha [S+\rho]\,, \\
  \partial_t \tilde{A}_{ij} &=& \beta^m \partial_m \tilde{A}_{ij}
        + 2\tilde{A}_{m(i} \partial_{j)} \beta^m
        - \frac{2}{3} \tilde{A}_{ij} \partial_m \beta^m
        + \alpha K\tilde{A}_{ij} \nonumber \\
     && - 2\alpha \tilde{A}_{im} \tilde{A}^m{}_j
        + \chi \left(
          \alpha \mathcal{R}_{ij} - D_i D_j \alpha
          - 8\pi \alpha S_{ij} \right)^{\rm TF}\,,\\
  \partial_t \tilde{\Gamma}^i &=& \beta^m \partial_m
          \tilde{\Gamma}^i
        + \frac{2}{3} \tilde{\Gamma}^i \partial_m \beta^m
        - \tilde{\Gamma}^m\partial_m \beta^i
        + \tilde{\gamma}^{mn} \partial_m \partial_n \beta^i
        \nonumber \\
     && + \frac{1}{3}\tilde{\gamma}^{im} \partial_m
          \partial_n \beta^n
        - \tilde{A}^{im} \left[
          3 \alpha \frac{\partial_m \chi}{\chi}
          + 2\partial_m \alpha \right]
        + 2\alpha \tilde{\Gamma}^i_{mn} \tilde{A}^{mn}
        \nonumber \\
     && - \frac{4}{3} \alpha \tilde{\gamma}^{im} \partial_m K
        - 16\pi \alpha j^i\,. \nonumber \\
        \label{eq:BSSNdtGamma}
\end{eqnarray}
Here, ``TF'' denotes the tracefree part and $\mathcal{R}_{ij}$ the
Ricci tensor associated with the physical three-metric $\gamma_{ij}$.
The promotion of auxiliary variables to independent status
introduces three additional constraints to the BSSN system given by
\begin{equation}
  \det \tilde{\gamma}_{ij} = 1\,,~~~~~
  \tilde{\gamma}^{mn} \tilde{A}_{mn} = 0\,,~~~~~
  \mathcal{G}^i \equiv \tilde{\Gamma}^i - \tilde{\gamma}^{mn}
        \tilde{\Gamma}^i_{mn} = 0\,.
  \label{eq:auxconstraintsBSSN}
\end{equation}
In practical applications, it turns out necessary for numerical stability
to control these auxiliary constraints in the following manner. (i) Enforce
$\tilde{\gamma}^{mn} \tilde{A}_{mn} = 0$ and (ii) either add the
constraint $\mathcal{G}^i$ to the right-hand side of Eq.~(\ref{eq:BSSNdtGamma})
\cite{Yo:2002bm} or, alternatively, substitute
on the right-hand side of Eq.~(\ref{eq:BSSNdtGamma}) all $\tilde{\Gamma}^i$
that appear in undifferentiated form
by their definition in terms of the metric $\tilde{\gamma}_{ij}$
\cite{Alcubierre:2000yz}.

Empirical studies quickly demonstrated that the BSSN system provides superior
numerical stability when compared with the ADM equations
(e.g.~\cite{Baumgarte:1998te}) and mathematical studies demonstrated
BSSN to provide a strongly hyperbolic formulation of the Einstein
equations \cite{Gundlach:2006tw}. The BSSN formulation is employed in
the binary BH breakthroughs by the Brownsville/Rochester and the NASA
Goddard groups \cite{Campanelli:2005dd,Baker:2005vv}.

The other formulation instrumental for the breakthroughs is based
on the Einstein equations in {\em harmonic gauge}
\cite{Einstein:1916cc} defined by the spacetime
coordinates satisfying the condition $\Box x^{\alpha} = -g^{\mu \nu}
\Gamma^{\alpha}_{\mu \nu} = 0$. In this form, the Ricci tensor
takes on the form
\begin{equation}
  R_{\alpha \beta} = -\frac{1}{2} g^{\mu \nu} \partial_{\mu}
        \partial_{\nu} g_{\alpha \beta} + \ldots\,,
\end{equation}
where the dots denote terms containing at most first derivatives of the
spacetime metric. In this form, the principal part of the Einstein equations
is that of the scalar wave operator and the equations are symmetric
hyperbolic. Harmonic coordinates have been used in the first
proofs of the local uniqueness of the Cauchy problem in GR
\cite{FouresBruhat:1952zz,Bruhat1962,Fischer1972}. A generalization
of this particularly appealing form of the Einstein equations to
arbitrary gauge is realized by promoting the functions
\begin{equation}
  H^{\alpha} \equiv \Box x^{\alpha} = -g^{\mu \nu} \Gamma^{\alpha}_{\mu \nu}\,,
  \label{eq:GHGH}
\end{equation}
to the status of independently evolved variables
\cite{Friedrich1985,Garfinkle:2001ni}. The resulting
system is often referred to as the {\em Generalized Harmonic Gauge} (GHG)
formulation and considers the generalized set of equations
\begin{equation}
  R_{\alpha \beta} -\nabla_{(\alpha} \mathcal{C}_{\beta)}
        = 8\pi \left( T_{\alpha \beta} - \frac{1}{2}Tg_{\alpha \beta}
          \right)\,,
  \label{eq:GHG_modEinstein}
\end{equation}
with the auxiliary constraints
$\mathcal{C}^{\alpha} \equiv H^{\alpha} - \Box x^{\alpha}$.
A solution to the Einstein equations is obtained by solving
Eq.~(\ref{eq:GHG_modEinstein}) subject to the condition
$\mathcal{C}^{\alpha}=0$. In practice, this is conveniently achieved by
prescribing initial data for $g_{\alpha \beta}$ and $\partial_t g_{\alpha
\beta}$ and initializing the $H^{\alpha}$ through Eq.~(\ref{eq:GHGH}).
If the initial data furthermore satisfy the Hamiltonian
and momentum constraints (\ref{eq:Ham}), (\ref{eq:mom}),
this can be shown to imply $\partial_t \mathcal{C}^{\alpha}=0$.
The Bianchi identities then ensure that the auxiliary constraint is
preserved under time evolution so that $\mathcal{C}^{\alpha}=0$
at all times in the continuum limit. For controlling
violations of these constraints at the discretized level in
numerical evolutions, Gundlach {\em et al.} \cite{Gundlach:2005eh}
suggested the addition of constraint damping terms which turned
out crucial in achieving the numerical stability required for binary
BH simulations \cite{Pretorius:2005gq}. With these terms,
the generalized Einstein equations (\ref{eq:GHG_modEinstein})
can be written in the form
\begin{eqnarray}
  g^{\mu \nu} \partial_{\mu} \partial_{\nu} g_{\alpha \beta} &=&
        - 2\partial_{\nu} g_{\mu (\alpha}\,\partial_{\beta)} g^{\mu \nu}
        - 2\partial_{(\alpha} H_{\beta)}
        + 2H_{\mu} \Gamma^{\mu}_{\alpha \beta}
        - 2\Gamma^{\mu}_{\nu \alpha} \Gamma^{\nu}_{\mu \beta} \nonumber \\
     && - 8\pi T_{\alpha \beta}
        + 4\pi T g_{\alpha \beta}
        - 2\kappa \left[2n_{(\alpha}\mathcal{C}_{\beta)}
          - \lambda g_{\alpha \beta} n^{\mu} \mathcal{C}_{\mu} \right]\,.
        \label{eq:GHG}
\end{eqnarray}
Here, $\kappa$ and $\lambda$ are user-specified constant parameters
which control the constraint damping.
The GHG formulation has been
used in Pretorius' breakthrough simulations \cite{Pretorius:2005gq};
see also \cite{Pretorius:2004jg}. \\[5pt]
%

\subsection{Gauge conditions}
In the previous section, we have seen that some of the evolution
variables are not determined by the Einstein equations. The lapse
function $\alpha$ and the shift vector $\beta^i$ are freely
specifiable in the ADM system (\ref{eq:dtgamma})-(\ref{eq:mom})
or the BSSN equations (\ref{eq:BSSNdtchi})-(\ref{eq:BSSNdtGamma})
and the $H^{\alpha}$ are undetermined in the GHG formulation
(\ref{eq:GHG}). Instead, these functions represent the coordinate
or gauge freedom of general relativity, and their choice leaves the
physical properties of the spacetime invariant. As one might expect
from this shared property, the two sets of gauge functions
are related; the normal component and spatial projection of
the $H^{\alpha}$ can be expressed in terms of the lapse $\alpha$
and shift $\beta^i$ respectively as given in Eqs.~(18), (19)
of Ref.~\cite{Pretorius:2004jg}. The geometrical meaning of
the gauge functions is more intuitively encoded in lapse and
shift and most investigations into the numerical properties
of different gauge choices have been carried out in terms of
these variables.

The simplest choice would appear to be given by $\alpha=1$
and $\beta^i=0$, referred to as {\em geodesic slicing with vanishing
shift}. The problems of using this gauge in numerical simulations,
however, have been demonstrated as early as 1978 by Smarr \& York's
\cite{Smarr:1977uf} time evolutions of the time symmetric slice
of the Kruskal-Schwarzschild spacetime. Setting $\alpha=1$ and
$\beta=0$, the numerically constructed hypersurfaces encounter
the BH singularity after an evolution time $t=\pi~M$;
cf.~the upper panel of their Fig.~2.
This behaviour highlights one key requirement to be met by
any numerically suitable set of gauge conditions: The evolution
in proper time should be slowed down in regions where the hypersurfaces
approach a singularity. This feature is commonly referred to as
{\em singularity avoiding slicing} and has been suggested first
in the form of {\em maximal slicing} $K=0$ \cite{Estabrook:1973ue}.
Singularity avoidance is achieved by letting the
lapse function vary in space and time such that it drops towards
zero in the vicinity of spacetime singularities.
For an illustration of this effect, we refer again to Fig.~2
in \cite{Smarr:1977uf} which contrasts maximal with geodesic slicing.
A wider class of singularity avoiding
slicings has been studied in the form of the Bona-Mass{\'o}
family \cite{Bona:1994dr} which includes maximal slicing as
a special case; see also \cite{Alcubierre:2002iq} and, for the
case of harmonic coordinates, \cite{Garfinkle:2001ni}.
It has been noticed, however, that due to the different advance in proper
time in different regions of the spacetime during the numerical
evolution, neighbouring grid points of a computational domain
may correspond to increasingly distant points in the spacetime. This
phenomenon, often referred to as {\em grid} or {\em slice stretching},
needs to be cured by a ``suitable shifting of grid points'' through the
use of a non-zero shift vector; see e.g.~\cite{Alcubierre:2002kk}.

Geometrically motivated shift conditions were used in the already
mentioned work by Smarr \& York \cite{Smarr:1977uf} in the form
of the {\em minimal distortion gauge}. Considering for example a small
sphere on a given hypersurface $\Sigma_t$, it can be shown that
the minimal distortion gauge preserves the spherical shape at leading
order whereas in general the shape will be sheared into an ellipse.
The maximal slicing condition furthermore preserves the volume of
the sphere. The numerical implementation of this shift is complicated
by the necessity to solve elliptic equations for the $\beta^i$;
for details see Sec.~4 and, in particular Eq.~(4.11) in
\cite{Smarr:1977uf}. In practice, it is much simpler to evolve the
shift in time according to parabolic or hyperbolic differential
equations which can be achieved with so-called ``driver'' conditions
\cite{Balakrishna:1996fe}.
Alcubierre {\em et al.} \cite{Alcubierre:2001vm}
have obtained such equations for the shift by relating
$\partial_t \beta^i$ or $\partial_t^2 \beta^i$ to the
elliptic operator obtained from the ``Gamma freezing'' condition
$\partial_t \tilde{\Gamma}^i=0$ where $\tilde{\Gamma}$ is the BSSN
variable defined in Eq.~(\ref{eq:BSSNvars}). They thus arrive at
the hyperbolic ``$\Gamma$-driver'' condition $\partial^2_t \beta^i
= \zeta \partial_t \tilde{\Gamma}^i -\xi \partial_t \beta^i$,
where $\zeta$ and $\xi$ are specifiable positive functions.
In a similar way, the Bona Mass{\'o} family replaces the
elliptic maximal slicing condition $K=0$ with an ``easier to
implement'' hyperbolic condition $\partial_t \alpha -\alpha^2
f(\alpha) [K-K(t=0)]$ where $f(\alpha)$ is a positive function.
By using a specific version of this slicing and $\Gamma$-driver
shift, Alcubierre {\em et al.} \cite{Alcubierre:2001vm}
managed to extract GWs from the
evolution of a distorted BH and drive the coordinates to a frame
where the system remains almost static at late times. A specific
version of the Bona-Mass{\'o} family is the so-called ``1+log''
slicing which sets $f(\alpha)=2/\alpha$ and enabled
Alcubierre {\em et al.} \cite{Alcubierre:2002kk} to evolve
BH data for long times above $1\,000~M$.

The breakthrough simulations of \cite{Campanelli:2005dd,Baker:2005vv}
obtained with the BSSN formulation employ variants
of the 1+log slicing and the $\Gamma$-driver shift condition
given by
\begin{eqnarray}
  \partial_t \alpha &=& \beta^m \partial_m -2\alpha K\,,
  \label{eq:dt_alpha} \\
  \partial_t \beta^i &=& \beta^m \partial_m \beta^i + \frac{3}{4} B^i\,, \\
  \partial_t B^i &=& \beta^m \partial_m B^i + \partial_t
        \tilde{\Gamma}^i - \eta B^i\,,
  \label{eq:dt_B}
\end{eqnarray}
or some minor modification of these equations; cf.~\cite{vanMeter:2006vi}.
Here, $\eta$ is a user specified constant or function of the coordinates.

The GHG formulation is motivated by the beneficial properties
of the Einstein equations in harmonic gauge, but stable numerical
evolutions of binary BH spacetimes have so far required at least
some component of the $H^{\alpha}$ to be non-zero.
Pretorius \cite{Pretorius:2005gq} sets $H_i=0$ and evolves the $t$ component
according to
\begin{equation}
  \Box H_t = -\xi_1 \frac{\alpha-1}{\alpha^{\eta}}
        + \xi_2 n^{\mu} \partial_{\mu} H_t\,,
  \label{eq:GHGgauge}
\end{equation}
where $\xi_1=19/m$, $\xi_2=2.5/m$, $\eta=5$ and $m$ denotes the mass
of one of the two (equal-mass) BHs. This choice prevents the lapse
from deviating too much from unity which may have caused instabilities
in earlier GHG evolutions \cite{Pretorius:2007nq}. For some further
discussion of gauge choices in the GHG formulation,
see e.g.~\cite{Pretorius2006,Szilagyi:2009qz}. \\[10pt]
%

\subsection{Boundary conditions and singularity treatment}
\label{sec:BCs}
Astrophysical BHs are commonly modeled as asymptotically flat spacetimes,
i.e.~the spacetime approaches the Minkowski limit far away from the
BH regions of strong curvature. Strictly speaking, this is an approximation
to the cosmological spacetimes that describe our universe, but for
most practical applications, as for example the modeling of GW signals
expected to be observed with laser interferometric detectors, it is
sufficient to include cosmological effects in the form of a redshift
factor $1+z$ multiplying the source mass. Asymptotically flat
spacetimes are of infinite extent and the challenge in numerical
relativity is to describe these inside compact computational domains.
The most elegant way to achieve this goal is to compactify the
spacetime coordinates and cover dimensions of infinite extent with
a finite coordinate interval as for example using maps of the type
$r\in [0,\infty)~\rightarrow~x\equiv 1/(r+1) \in (0,1]$. Applied to
3+1 splits of the spacetime, however, this often results in an asymptotically
infinite blue shifting of gravitational radiation; the wavelength
of the radiation asymptotically shrinks to zero as measured in the
compactified coordinate and fails to be resolved numerically.
Characteristic formulations of the Einstein equations
\cite{Winicour:2005ge}, on the other
hand, are ideally suited for such a treatment, as the coordinates
are constructed in terms of light cones, GW signals have constant
phase along the characteristic coordinate curves and no resolution
problems arise. Here lies one of the attractive features of
characteristic formulations mentioned above in Sec.~\ref{sec:Einsteinian}.
Inside
3+1 formulations, such behaviour can be obtained by slicing the
spacetime with hypersurfaces that are spacelike everywhere, but
become asymptotically null at infinity. This type of slicing can
be obtained for example using {\em hyperbolic slicing}
$K=\mathrm{const} \ne 0$ and plays an important role in the so-called
{\em conformal field equations} (see \cite{Frauendiener:2000mk} and
references therein), but has, to our knowledge, not yet been
applied successfully to simulate BH binaries.

In practice, most numerical applications model only a finite subset of
the total spacetime and impose boundary conditions at large but finite
distance from the BHs. Ideally,
such boundary conditions satisfy the following
three requirements. (i) Ensure well posedness of the IBVP, (ii)
compatibility with the Einstein constraint equations, and (iii)
a correct representation of the physical boundary conditions, typically
minimization of the ingoing gravitational radiation \cite{Rinne:2006vv}.
Such conditions have been studied mostly
for the GHG formulation; see \cite{Babiuc:2006ik,Rinne:2007ui,Ruiz:2007hg}
and references therein.

Boundary conditions meeting the above criteria have not yet been developed
for the BSSN formulation\footnote{But see
\cite{Hilditch:2012fp} for investigations
using a modification of BSSN commonly referred to as the conformal $Z4$ system.}
and numerical applications of the BSSN system therefore resort to
an approximation using {\em outgoing radiation} or {\em Sommerfeld}
conditions. The assumption underlying this approach is that the evolution
variables $f$ approach a constant background value $f_0$ far away from
the strong-field sources and deviations from this value at finite
radius $r$ can be written as $f=f_0 + u(t-r) /r^n$ with a positive, integer
$n$. The outgoing radiation condition $\partial_t u + \partial_r u=0$
for the radiative deviations then translates into the boundary condition
\cite{Alcubierre:2002kk}
\begin{equation}
  \partial_t f + n \frac{f-f_0}{r} + \frac{x^i}{r} \partial_i f =0\,,
\end{equation}
where $x^i$ denote Cartesian coordinates
and $r^2= \sum_i (x^i)^2$. These conditions
are not without problems: (i) The system is over-determined because
the number of conditions imposed exceeds that of the ingoing characteristics;
(ii) Sommerfeld conditions are not constraint satisfying, and (iii)
the non-exact outgoing nature of these conditions at finite radii may
introduce spurious reflections. In spite of these caveats, Sommerfeld
conditions turn out to work rather well and robustly in many numerical
applications (see e.g.~\cite{Rinne:2007ui}) and are the method of choice
for the moving puncture breakthroughs of the Brownsville/RIT and Goddard
groups
\cite{Baker:2005vv,Campanelli:2005dd}. Pretorius \cite{Pretorius:2005gq},
instead, uses a compactified domain and overcomes the problem
of under-resolving the blue-shifted radiation by damping the radiation
through numerical viscosity and thus effectively emulates
no-ingoing-radiation conditions.

A second type of boundary conditions arises in NR
applications through the presence of the spacetime singularities.
These singularities typically manifest themselves in the form
of diverging or vanishing metric components as for example the
$g_{rr} = (1-2M/r)^{-1}$ in the Schwarzschild metric in Schwarzschild
coordinates. Computer simulations react with non-assigned numbers
to the resulting
infinities which rapidly swamp the entire computational domain and
render the simulation practically useless. One elegant approach
to handle this problem employs so-called ``puncture'' initial
data (see Sec.~\ref{sec:Inidata})
and factors out the singular part of the BH data throughout
the time evolution; see e.g.~\cite{Anninos:1995am,Alcubierre:2002kk}
for details of these {\em fixed puncture} evolutions.
In this approach, the BHs remain at fixed coordinate location
throughout the evolution and it appears to be difficult to
construct long-term stable coordinate conditions for BH inspiral in
this approach; see \cite{Bruegmann:2003aw} for the most advanced
application of this type leading to about one orbit of BH inspiral.

An alternative method to handle singularities which has become popular over
the years is the {\em BH} or {\em singularity excision} technique
attributed to Unruh \cite{Thornburg1987}. By virtue of Penrose's
cosmic censorship conjecture (see e.g.~\cite{Wald1997}),
spacetime singularities should be cloaked inside an event horizon
such that the spacetime exterior to the horizon is causally
disconnected from events inside the horizon. In particular, the
exterior spacetime should not be affected by completely removing
a finite region around the singularity from the numerical evolution
as long as the excised region remains inside the event horizon
or, as usually done in practice, is located inside the
apparent horizon (AH) \cite{Thornburg:2006zb} on each hypersurface $\Sigma_t$.
\begin{figure}[t]
  \centering
  \includegraphics[height=180pt,clip=true,angle=0]{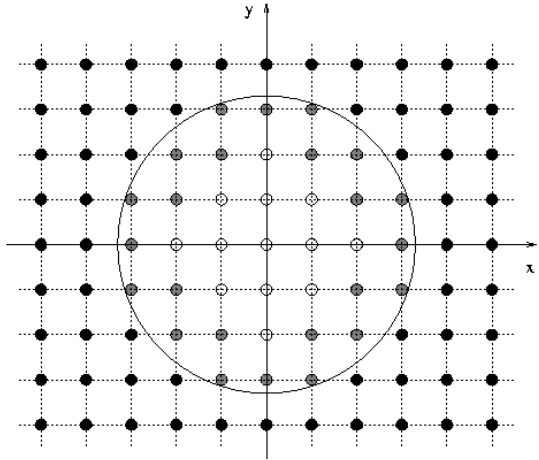}
  \caption{Illustration of BH excision with one
           spatial dimension suppressed. Black grid points are updated
           regularly in time, white points inside the AH
           (large circle) are excluded from the time evolution and
           gray points mark the excision boundary and need to be updated
           in time using sideways differencing operators
           \cite{Pretorius:2004jg}, extrapolation
           \cite{Shoemaker:2003td} or are filled in through regular
           update with spectral methods \cite{Scheel:2006gg}.
          }
  \label{fig:excision}
\end{figure}
This is illustrated in Fig.~\ref{fig:excision} where the large circle
represents the AH and the region consisting of the
white (empty) small circles is removed from the computational
domain whereas black (filled) points are updated regularly. The update
in time at a given grid point requires data from neighbouring
points to evaluate spatial derivatives, and therefore the excision boundary
(gray circles) may require some special treatment. This can be achieved
either by sideways differencing operators \cite{Pretorius:2004jg},
extrapolation from data on grid points further out
\cite{Shoemaker:2003td} or calculating the function values from
the spectral expansion in spectral codes \cite{Scheel:2006gg}.
The first of these schemes is the method employed in Pretorius'
work \cite{Pretorius:2005gq}. Rather astonishingly, the
moving puncture method \cite{Campanelli:2005dd, Baker:2005vv}
does not implement an explicit excision scheme but instead uses
finite differencing stencils right across the BH singularities.
The surprising success of this method has been explored in more
depth in \cite{Hannam:2008sg,Brown:2009ki,Dennison:2010wd}
and references therein. The singularity of puncture type initial
data is a coordinate singularity that contains spatial infinity
of the far side of the wormhole geometry compactified into a
single point. In moving puncture evolutions, however, these initial
data rapidly change from a wormhole to a so-called ``trumpet''
geometry which is only partially covered by the computational domain
because of the discrete structure of the numerical grid;
cf.~Fig.~1 in \cite{Brown:2009ki}. The singularity, instead,
``falls through the grid'' and the moving puncture technique can therefore
be interpreted as an indirect excision method provided by the finite
grid resolution. \\[10pt]
%

\subsection{Discretization and mesh refinement}
Computers operate with finite arrays of numbers or, strictly speaking, with
binary numbers that are readily converted
into integers in the decimal system (exactly) or floating point numbers
(with finite precision, the so-called ``round-off error'').
In the numerical calculation of solutions to
differential equations there thus arises the challenge to describe
functions and their derivatives in terms of finite arrays of numbers.
This process is commonly referred to as ``discretization'' and most
commonly achieved in computational analysis using one of four
methods, (1) finite differencing, (2) spectral methods, (3)
finite elements or (4) finite volume methods. The latter two have,
to our knowledge, not yet been applied to NR
simulations of BH binaries.

Spectral methods operate with an expansion of the physical variables
in basis functions and facilitate exceptionally efficient and
accurate numerical modeling. In particular, they result in
exponential convergence when applied to problems with smooth solutions.
This exponential convergence would be spoiled by functions containing
singularities as present in BH spacetimes, but this drawback can
be overcome by removing the singular points from the computational domain
through BH excision. The high accuracy of spectral methods has been
brought to fruit in BH binary evolutions with the SpEC code
\cite{Boyle:2007ft,Lovelace:2011nu} and in the constraint solving
for the construction of initial data \cite{Ansorg:2004ds,Zilhao:2011yc};
for a review of spectral methods in
NR see \cite{Grandclement:2007sb}.

In finite differencing the computational domain consists of one
or more discrete {\em grids} (cf.~Fig.~\ref{fig:excision}),
functions are represented by their values at the grid points
and derivatives are approximated through Taylor expansion
by differences of the function
values on neighbouring grid points. The accuracy of this
approximation depends on the number of neighbouring points
used and is typically measured in terms of the leading order
term of a Taylor expansion in
the grid spacing $\Delta x$ between grid points;
for an example of the finite differencing expressions thus
obtained see for example Sec.~2 in \cite{Zlochower2005}.
\begin{figure}[t]
  \centering
  \includegraphics[height=180pt,clip=true,angle=-0]{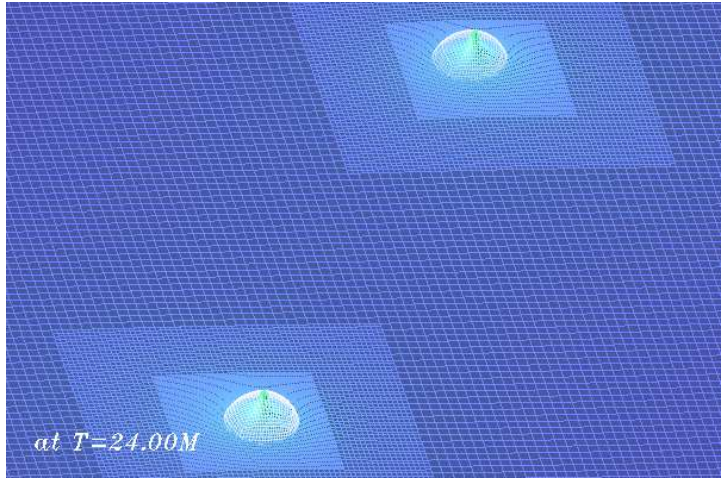}
  \caption{Illustration of mesh refinement with a moving boxes approach.
           In this example, two boxes are centered around either BH
           (marked by the spherical horizon surfaces) and immersed inside
           one large common grid.}
  \label{fig:mesh_refinement}
\end{figure}
The main advantage of finite differencing methods is their
comparative simplicity and high robustness in modeling a wide
class of extreme BH binary systems with little if any modifications
in the methodology; see e.g.~\cite{Lousto:2010ut,Sperhake:2012me}.

As mentioned in Sec.~\ref{sec:Einsteinian}, BH spacetimes involve
a wide range of length scales which cannot be efficiently accommodated
inside uniform grids and thus require the use of {\em mesh refinement},
i.e.~a grid resolution that varies in space and time. BH horizons are
remarkably rigid objects and typically maintain an approximately
spherical shape throughout inspiral and even during the merger phase,
so that high accuracy can be achieved by an approach sometimes referred
to as ``moving boxes''. The computational domain
consists of a set of nested boxes centered around the individual BHs
immersed inside one or more large boxes containing both BHs.
The grid spacing $\Delta x$ increases (typically by a factor $2$) from
each box to the next outer one; for a graphical illustration of this
method see Fig.~\ref{fig:mesh_refinement}. More general shapes
of the different {\em refinement levels} can be achieved by arranging
a larger number of boxes in a ``lego-style'' manner or by using
threshold values on physical variables, as for example the
curvature scalar, as a criterion to introduce new grid points.
Communication between
the different levels of a grid hierarchy is achieved by some
form of interpolation, typically between a given level and its
two neighbours in the hierarchy. Mesh refinement was introduced to
NR by Choptuik's seminal study
on critical collapse in spherical symmetry \cite{Choptuik:1992jv}
and first applied to BHs in three spatial dimensions by Br{\"u}gmann
\cite{Bruegmann:1996kz}. Mesh refinement of this type has been implemented
in Pretorius' code in the form of the {\sc Pamr/Amrd}
\cite{PAMRAMRD} package while the Goddard group has used
{\sc Paramesh} \cite{MacNeice2000}. In contrast, the Brownsville/RIT
group achieved position dependent resolution through the use
of a transformation from standard Cartesian coordinates to so-called
``fish-eye'' coordinates using logarithms and hyperbolic functions
such that the spacing between grid points increases away from the
strong curvature region near the origin
\cite{Alcubierre:2002kk,Baker:2001sf,Zilhao:2013dta}.

Further packages used for mesh refinement include
\textsc{Bam}~\cite{Bruegmann:1996kz},
\textsc{Had}~\cite{HAD},
\textsc{Samrai}~\cite{Samraiweb} and
\textsc{Carpet}~\cite{Schnetter:2003rb, Carpetweb}.
The latter is
provided as part of the {\sc Cactus} \cite{Goodale2002} and
{\sc Einstein Toolkits} \cite{EinsteinToolkit} which are publically
available environments used by various NR groups.
In spectral applications, a similar method to accommodate vastly different
length scales is implemented in the form of subdomains of varying shapes
that communicate through matching conditions at the boundary for touching
domains or in small regions of overlap;
see e.g.~\cite{Pfeiffer:2002wt,Buchman:2012dw}. \\[10pt]
%

\subsection{Initial data}
\label{sec:Inidata}
The task in generating initial data in numerical
relativity is two-fold. (i) The data must satisfy the Einstein
constraint equations (\ref{eq:Ham}), (\ref{eq:mom}) and (ii) they
need to represent a realistic snapshot of the physical system under study.
A natural starting point for the construction of initial data is to apply
modifications to existing analytic BH solutions. An
analytic solution of particular relevance for this purpose is
the Schwarzschild solution in isotropic coordinates
\begin{equation}
  ds^2 = -\left( \frac{2r-M}{2r+M} \right)^2 dt^2
        + \left( 1+ \frac{M}{2r} \right)^4 \left( dr^2 + r^2 d\theta^2
        + r^2 \sin^2 \theta\,d\phi^2 \right)\,.
  \label{eq:isotropic}
\end{equation}
Its scope for generalization
becomes clear in a systematic approach to solving
the Einstein constraints based on the York-Lichnerowicz split
\cite{Lichnerowicz1944,York:1971hw,York:1972sj}. This method consists
of a conformal transformation of the spatial metric $\gamma_{ij}
= \psi^4 \bar{\gamma}_{ij}$ and a
{\em conformal traceless split} of the extrinsic curvature
\begin{equation}
  K_{ij} = A_{ij} + \frac{1}{3} \gamma_{ij} K\,,~~~~~
  A^{ij} = \psi^{-10} \bar{A}^{ij}\,,~~~~~
  A_{ij} = \psi^{-2} \bar{A}_{ij}\,.
  \label{eq:ctsplit}
\end{equation}
By further decomposing $\bar{A}_{ij}$ into a longitudinal piece
plus a transverse traceless part, the momentum constraints
simplify considerably \cite{Cook:2000vr}.
Similar simplifications are achieved by using instead of
Eq.~(\ref{eq:ctsplit}) a {\em physical transverse traceless split}
or the so-called {\em thin-sandwich decomposition}; for details see
\cite{York1999,Cook:2000vr,Caudill:2006hw} and references therein.

The Hamiltonian and momentum constraints take on a particularly
simple form if one further requires that (i) the trace of the extrinsic
curvature vanishes $K=0$, (ii) the conformal metric is flat
$\bar{\gamma}_{ij} = f_{ij}$ where $f_{ij}$ is the Euclidean metric
(not necessarily in Cartesian coordinates), and (iii) the spatial
metric is asymptotically flat, $\lim_{r \rightarrow \infty} \psi = 1$.
Then the momentum constraints decouple from the Hamiltonian constraint
and form a set of equations for the $\bar{A}_{ij}$. The simplest solution
for these equations is the trivial one $\bar{A}_{ij}=0$ in which case
the Hamiltonian constraint becomes
$\bar{\Delta} \psi=0$, where $\bar{\Delta}$ is the Laplace operator
associated with the flat metric $f_{ij}$.
This Laplace equation for the conformal
factor $\psi$ is solved by the spatial part of
the Schwarzschild metric (\ref{eq:isotropic}) $\gamma_{ij}=\psi^4
f_{ij}$ with $\psi= 1+M/(2r)$. By linearity of the Laplace equation,
we can obtain initial data containing multiple BHs by
superposing $N$ solutions of this type according to
$\psi = 1 + \sum_{A=1}^N m_A/(2 |\vec{r} - \vec{r}_A|)$, where
$m_A$ and $\vec{r}_A$ denote mass and location of the BHs.
This solution is known as Brill-Lindquist \cite{Brill:1963yv}
data and represents a snapshot
of $N$ BHs at the moment of time symmetry.
A similar type of data differing only in the symmetry conditions at
the throat of the worm hole(s)
has been constructed by Misner \cite{Misner:1960zz}. Both, Brill-Lindquist
and Misner data formed the starting point of many BH evolutions over the
previous decades.

Quite remarkably, under the assumption of conformal flatness and $K=0$, the
momentum constraints even admit non-vanishing analytic solutions for the
extrinsic curvature, the
Bowen-York \cite{Bowen:1980yu} data
\begin{equation}
  \bar{A}_{ij} = \frac{3}{2r^2} \left[ P_i n_j + P_j n_i - (f_{ij}
        - n_i n_j) P^kn_k \right] + \frac{3}{r^3}
        \left( \epsilon_{kil} S^l n^k n_j
        + \epsilon_{kjl} S^l n^k n_i \right)\,,
  \label{eq:BowenYork}
\end{equation}
where $r$ is the areal radius associated with the flat metric
$f_{ij}$, $n^i$ the unit, outgoing radial vector and $P^i$, $S^i$
are free parameters that correspond to the total linear and angular
momentum of the initial hypersurface \cite{York1980}.
The momentum constraints are linear in $\bar{A}_{ij}$, so that multiple
solutions of the type (\ref{eq:BowenYork}) can be superposed. Equation
(\ref{eq:BowenYork}) gives the generalization of Brill-Lindquist data
for non-zero BH momenta. For generalization of Misner data, one needs
to construct inversion-symmetric data of the type (\ref{eq:BowenYork})
using the method of images (see Sec.~3.2.1 in \cite{Cook:2000vr}).

In the conformal-flatness approximation with $K=0$ and non-vanishing Bowen-York
extrinsic curvature (\ref{eq:BowenYork}), the Hamiltonian constraint
becomes
\begin{equation}
  \bar{\Delta} \psi + \frac{1}{8} K^{mn} K_{mn} \psi^{-7} = 0\,.
  \label{eq:CFHam}
\end{equation}
This elliptic equation is often solved by decomposing the conformal
factor into a Brill-Lindquist piece $\psi_{\rm BL} = 1+M/(2r)$ plus
a regular contribution $u$. Under such decomposition,
the existence and uniqueness of $C^2$ regular
solutions $u$ to Eq.~(\ref{eq:CFHam}) has been proven by Brandt and Br{\"u}gmann
\cite{Brandt:1997tf} and the data thus constructed are commonly
referred to as {\em puncture data}. The Schwarzschild solution
(\ref{eq:isotropic}) in isotropic coordinates is the simplest non-trivial
solution of this type, a single BH with zero linear and angular momentum.
Alternative to the puncture method, the initial data formalism
summarized here has also been applied, in some flavor or other,
to the construction of BH excision data;
see e.g.~\cite{Pfeiffer:2002wt,Ansorg:2006gd,Grandclement:2007sb}.

In spite of their popularity in time evolutions, the conformal flatness
nature of puncture data results in some restrictions. In particular,
the Kerr \cite{Kerr1963} spacetime describing a single rotating
BH does not contain a maximal, conformally flat hypersurface
\cite{Garat2000,ValienteKroon:2003ux}. Puncture data with non-zero
Bowen-York angular momentum $S^i$ therefore not only contain
a rotating BH but some further gravitational fields which
manifest themselves as pulses of {\em spurious
radiation} colloquially referred to as ``junk radiation''. While this
spurious GW pulse is often small, it increases non-linearly with
the Bowen-York parameters $P^i$ and $S^i$. In particular, this imposes
a practical limit of the initial dimensionless spin parameter of
BH configurations of $\approx 0.93$ \cite{Cook1989,Dain:2002ee}
and has motivated the construction of initial data without the
conformal-flatness assumption by either applying most of the
puncture formalism to a non-conformally
flat background metric \cite{Krivan:1998td,Hannam:2006zt}
or applying the conformal-thin-sandwich
method to a background of superposed Kerr-Schild data
\cite{Lovelace:2008tw,Lovelace:2010ne}.

Puncture data have been the starting point for the majority of BH binary
simulations in the past decade and, as suggested by the name, were also
used in the moving puncture breakthroughs
\cite{Campanelli:2005dd,Baker:2005vv}. A conceptually rather different
approach was used in Pretorius' simulations. Instead of using
initial data containing BHs, the simulations start with matter
in the form of scalar field clouds concentrated and boosted such that
they rapidly collapse into a BH with velocity corresponding
approximately to a binary configuration in quasi-circular orbit
\cite{Pretorius:2005gq}. \\[10pt]
%

\subsection{Diagnostics}
The extraction of physical information from a
BH binary simulation is typically a diagnostic process that uses
the numerically constructed fields but has at most minor
impact\footnote{For example, the calculation of an apparent horizon
may be used for choosing regions for BH excision or curvature
quantities may be used as criteria for mesh refinement.} on
the actual time integration of the fields. Their significance
for BH binaries therefore mostly consists in understanding the results
rather than solving the two-body problem itself and we shall only
briefly summarize here the most important diagnostic quantities
but provide references for more details. The diagnostic quantities can
be loosely classified into three groups: properties of the global
spacetimes, the GW signal and the BH horizons.
\\[10pt]
\noindent
{\bf \em Global quantities:} For asymptotically flat spacetimes,
the total mass-energy and the linear
momentum of a spacetime is given by the ADM
mass and momentum \cite{Arnowitt:1962hi}. If the coordinate system
is chosen such that in the limit of infinite separation $r$ from the
strong-field sources the spacetime metric deviates from the
Minkowski metric $\eta_{\mu \nu}$ according to
$g_{\mu \nu} = \eta_{\mu \nu} + \mathcal{O}(1/r)$, the ADM mass and
momentum is given in terms of the ADM variables by the integrals
\begin{eqnarray}
  M &=& \frac{1}{16\pi} \lim_{r\rightarrow \infty}
        \oint_{S_r} \delta^{mn}( \partial_n \gamma_{mk}
        - \partial_k \gamma_{mn}) \hat{r}^k\,dS\,, \\[10pt]
  P_i &=& \frac{1}{8\pi} \lim_{r\rightarrow \infty} \oint_{S_r}
        (K_{mi}-\delta_{mi}K) \hat{r}^m\,dS\,,
\end{eqnarray}
where the components $\gamma_{mn}$, $K_{mn}$ are in Cartesian coordinates,
$\hat{r}^i=x^i/r$ is the outgoing unit normal to the surface of integration,
$S_r$ denotes the 2 sphere of coordinate radius $r$ and $dS$ is the
standard surface element of the 2 sphere. Under a more restricted
class of gauge conditions (see \cite{Gourgoulhon:2007ue} for details), one
can also calculate the angular momentum of the spacetime from
\begin{equation}
  J_i = \frac{1}{8\pi} \lim_{r\rightarrow \infty}
        \oint_{S_r} (K_{jk} - K \gamma_{jk}) \boldsymbol{\xi}_{(i)}^j
        \hat{r}k\,dS\,,
\end{equation}
where the $\boldsymbol{\xi}_{(i)}$ are the Killing vectors associated with
the asymptotic rotational symmetry. For an extended discussion of the
ADM variables, the reader is referred to Sec.~7 of \cite{Gourgoulhon:2007ue}
and references therein. \\[10pt]
\noindent
{\bf \em Horizons:}
Event horizons are a defining criterion of BH spacetimes and
mark the boundary between points from which null geodesics can
reach infinity and points from which they cannot. Even though
numerical algorithms have been developed for the calculation of
event horizons in BH spacetimes \cite{Hughes:1994ea,Diener:2003jc,Cohen:2008wa},
it is often more convenient to instead calculate the {\em apparent
horizon} \cite{Thornburg:2006zb}. An AH is defined
as the {\em outermost marginally trapped surface} on a spatial
slice $\Sigma_t$. This condition can be shown to result in
an elliptic equation for the outgoing normal direction $s^i$ of
the two-dimensional AH (see e.g.~\cite{Gundlach:1997us})
\begin{equation}
  q^{mn} D_m s_n -K +K_{mn} s^m s^n = 0\,,
\end{equation}
where $q_{mn}$ is the induced 2-metric  on the horizon surface.
Unlike an event horizon, the AH can be calculated independently
from the data on each hypersurface without further knowledge of the
spacetime. Under the assumption of cosmic censorship and certain
energy conditions, it can furthermore be shown that if a hypersurface
$\Sigma_t$ contains an AH, it will coincide or lie within the event
horizon's cross section with $\Sigma_t$ \cite{Hawking:1973uf,Wald1984}.
The {\em irreducible mass} of a BH is directly encoded in the AH
surface area by $M_{\rm irr}^2 = 16\pi A_{\rm AH}$ and, combined with
the BH spin $S$ gives the total BH mass through $M^2 = M_{\rm irr}^2
+S^2/(4M_{\rm irr}^2)$ \cite{Christodoulou:1970wf}. This formula also
provides an estimate for the dimensionless BH spin $j=S/M^2$ in terms
of the AH area and equatorial circumference $2\pi A_{\rm AH}/C_e^2
= 1+\sqrt{1-j^2}$; see e.g.~\cite{Sperhake:2009jz}. The importance
of horizons in the analysis of BH spacetimes follows to a large extent
from the {\em isolated} and {\em dynamic horizon} framework developed
by Ashtekar and coworkers \cite{Ashtekar:2004cn}. \\[10pt]
\noindent
{\bf \em Gravitational Waves:}
Arguably the most fundamental difference between the Newtonian and
the general relativistic two-body problem is the dissipative character of
the latter; energy and momenta of the binary are not conserved but
partly radiated away in the form of GWs.
Large-scale international efforts are dedicated to directly detect
GWs with ground-based
laser interferometric detectors LIGO, VIRGO, GEO600, KAGRA,
future space missions of LISA type or pulsar timing arrays
\cite{LIGOweb,advancedVIRGO,GEO600web,KAGRAweb,ELISAweb,IPTAweb}
and the theoretical prediction of the expected GW signals from astrophysical
sources has been one of the main motivations
of NR.

The most common approach to calculate GWs in BH simulations is
based on the Newman-Penrose formalism \cite{Newman:1961qr}
where the 10 independent components of the Weyl tensor are
projected onto a tetrad consisting of one outgoing and one ingoing
null vector
$\boldsymbol{\ell}$ and $\boldsymbol{k}$ as well as two complex spatial
null vectors $\boldsymbol{m}$ and $\bar{\boldsymbol{m}}$.
The interpretation of the resulting five complex scalars
$\Psi_n$, $n=0,\,\ldots,\,4$ in terms of GWs is based on
the work of Bondi, Sachs and Penrose and coworkers
\cite{Bondi:1958aj,Bondi:1962px,Sachs:1962wk,Penrose:1962ij} and application
of this formalism in NR requires a careful choice of the tetrad.
In particular, the tetrad must correspond to a Bondi frame which
can be realized, for example, by choosing a so-called
{\rm quasi-Kinnersley} tetrad
\cite{Beetle:2004wu,Zhang:2012ky}, i.e.~a tetrad that converges to
the Kinnersley tetrad \cite{Kinnersley:1969zza} as the spacetime
approaches Petrov type D. A particularly convenient choice is realized
in the so-called transverse frame where the outgoing gravitational
radiation is encoded in one complex scalar
(see e.g.~\cite{Nerozzi:2008ng})
\begin{equation}
  \Psi_4 = -C_{\alpha \beta \gamma \delta} \ell^{\alpha} \bar{m}^{\beta}
        \ell^{\gamma} \bar{m}^{\delta}\,.
\end{equation}
Even though $\Psi_4$ is well defined at infinity only, it is in practice
often extracted at large but finite radii and this procedure generates
various potential systematic errors which are discussed
in \cite{Lehner:2007ip}. The effect of these ambiguities is sometimes
mitigated by extrapolating results at different finite radii to
infinity \cite{Hinder:2013oqa} which suggests that the errors thus
obtained are of the order $\mathcal{O}(\%)$. The Newman-Penrose scalar
$\Psi_4$ is commonly decomposed into multipoles $\psi_{lm}$ according to
\begin{equation}
  \Psi_4(t,\theta,\phi) = \sum_{l,m} \psi_{lm}(t)\, Y^{-2}_{lm}(\theta,\phi)\,,
\end{equation}
where $Y^{-2}_{lm}$ are spherical harmonics of spin weight $-2$. Often,
the radiation is dominated by one or a few multipoles which can then
be displayed in the form of functions of time. The Newman-Penrose
scalar furthermore provides the energy, linear and angular momentum
carried by the GWs which are obtained from straightforward integrals
of $\Psi_4$ and its projections onto asymptotic Killing vectors
\cite{Ruiz:2007yx}.

Other methods for estimating the gravitational radiation have been applied
in NR. (i) Perturbative wave extraction is based on the
Regge-Wheeler-Zerilli-Moncrief formalism
\cite{Regge:1957td,Zerilli:1971wd,Moncrief:1974am} and constructs
master functions from the deviation of the spacetime metric from
a Schwarzschild background; see e.g.~\cite{Nagar:2005ea,Sperhake:2005uf}.
(ii) The Landau-Lifshitz pseudo tensor \cite{Landau:1980} is constructed
by mapping the curved, physical spacetime onto an auxiliary flat spacetime
with metric $\eta_{\mu \nu}$. This leads to expressions for the radiated
energy and momenta; for applications see e.g.~\cite{Lovelace:2009dg}.
(iii) In characteristic formulations of the Einstein equations, the
Bondi news function \cite{Bondi:1962px}
provides a direct measure of the GW signal which has also been used
in 3+1 NR through Cauchy-characteristic extraction
\cite{Reisswig:2009us, Babiuc:2010ze}. This method provides a
particularly accurate extraction since it is performed by construction
at infinity; for a comparison with other methods see \cite{Reisswig:2010cd}.

\section{A brief history of black-hole simulations}
\label{sec:History}

In this section we will briefly review the main developments in NR
leading to the breakthroughs of 2005. It is beyond the scope of this
work to present a comprehensive history of the enormous amount of
work and publications generated in this field over the last 50 years.
Our review should therefore be understood as a potentially biased
precis intended to give the reader a rough guideline of NR's history.
The articles quoted in this section contain many further references
the reader will find a valuable source for a more thorough account.

The earliest documented effort to generate BH spacetimes by numerically
solving Einstein's equations was done half a century ago by Hahn \& Lindquist
\cite{Hahn1964}. It is worth noting that at the time the notion of BHs,
horizons and the area theorems were not yet understood. Furthermore,
virtually nothing about the delicacies of all the ingredients discussed
in the previous section was known at the time.
It is thus not too surprising that they could evolve their data for very short
times only. And yet, their first steps into uncharted territory
demonstrated a genuinely new alternative for the modeling
of BHs even if few, at the time, would have predicted what kind
of avalanche of numerical explorations had been kicked loose.

Starting in the late 1960s, the problem was reinvestigated in an
effort initiated by DeWitt which led to PhD theses
by \v{C}ade{\v{z}}, Smarr and Eppley \cite{Cadez1971,Smarr1975,Eppley1975}.
These works implemented the ADM equations in axisymmetry
using a specific type of coordinates often referred to as
{\em \v{C}ade\v{z}} coordinates and thus evolved head-on
collisions starting with Misner data, testing several gauge
conditions including maximal slicing, vanishing shift and minimal
distortion shift. Their equal-mass head-on collisions predict a GW energy
of about $0.1~\%$ of the total mass, albeit with uncertainties
of a factor a few; for details see \cite{Smarr1976,Smarr:1977uf,Smarr1979}.
This value turned out to be correct within a factor of about 2.

The next burst of efforts took place in the 1990s, much of it
as part of the ``Binary Black Hole Grand Challenge
Project'' \cite{Choptuik:1997}. Earlier studies of this decade
still employed axisymmetry and similar techniques as above,
but on significantly improved
computational architecture. Anninos {\em et al.} \cite{Anninos:1993zj,
Anninos:1994gp,Anninos:1995vf}
extracted the $l=2$ and $l=4$ multipoles of the emitted GW signal,
calculated AHs and compared results obtained with different wave extraction
methods; they confirmed the earlier estimates for
the emitted GW energy within error bars and found good agreement with
close-limit \cite{Price:1994pm} predictions for small initial separations
of the BHs; see also \cite{Baker:1996bt} for collisions of boosted BHs.
One of the most memorable results of these axisymmetric
BH simulations is the ``pair-of-pants'' picture obtained when calculating
the BH horizons in a merger process; cf.~Fig.~10 in \cite{Matzner:1995ib}.
Using a special class of ``body fitting'' coordinates,
Anninos and Brandt \cite{Anninos:1998wt} performed the first
evolutions of unequal-mass BH head-on collisions. By extracting GW modes
up to $l=4$, they validated perturbative results in the close and large
separation limit.
Further axisymmetric studies include initially distorted,
rotating BHs and the resulting GW emission \cite{Brandt:1994ee}
and accretion onto rotating BHs \cite{Brandt:1998cv}.

The first fully 3+1 dimensional BH simulations were presented in 1995 by Anninos
{\em et al.} \cite{Anninos:1995am} with the so-called
``G-code''. This code is based on the ADM formulation, uses
Schwarzschild initial data in isotropic coordinates, different types
of singularity avoiding slicings and a shift that locks the coordinate
radius to the apparent horizon location. It produced numerically stable
solutions of a Schwarzschild BH
for up to $t \approx 50~M$.
The GW signal including
BH ringdown obtained with the G-code were found to agree well with those
from axisymmetric codes \cite{Camarda:1997qv,Camarda:1998wf}.
Further development of their 3+1 code enabled the Grand Challenge Alliance
to evolve a single BH that moves across the computational domain
with $0.1$ times the speed of light for a total time of about
$60~M$ \cite{Cook:1997na}. Around the same time,
mesh refinement was first used in 3+1 simulations of a BH by Br{\"u}gmann
\cite{Bruegmann:1996kz}. The year 1997 saw the release of
{\sc Cactus} 1.0 \cite{Cactusweb}, a freely available environment
for the development of
parallel, scalable, high-performance multidimensional component-based
code for NR and other numerical applications.

The first binary BH merger in 3+1 NR was simulated by Br{\"u}gmann
\cite{Bruegmann:1997uc} in grazing collisions using the ADM equations
and the fixed puncture technique.
The first grazing collisions
of BHs using excision were performed with {\sc Agave},
a revised version of the Grand Challenge code \cite{Brandt:2000yp}.

As the second millennium drew to a close,
however, it was still a general feature of the space-time-split based
codes, in axisymmetry or full 3+1, to be limited by numerical instabilities to
life times of the order of $\mathcal{O}(100~M)$. This is in sharp
contrast to the remarkable stability properties achieved at the
same time using characteristic methods which facilitated long-term
stable simulations of single distorted, moving or rotating BHs
with lifetimes up to $60\,000~M$
\cite{Lehner:1998ti,Gomez:1998uj}. As mentioned in
Sec.~\ref{sec:Einsteinian}, characteristic
codes have not yet been generalized successfully to binary BH
spacetimes where the formation of caustics and the ensuing
breakdown of the null coordinates have so far represented an insurmountable
obstacle.
A different approach named ``Lazarus'' \cite{Baker:2000zh,Baker:2001sf}
was developed around the turn of the millennium in order to maximize
the scientific output obtained from 3+1 evolutions as available at the time.
In this eclectic approach the relatively short numerical simulations
are matched to perturbative calculations once a merger into a single BH
has occurred and the spacetime is perturbatively close to a Kerr BH.
The preceding inspiral phase, instead, is to be described by PN
methods which provide initial data for the numerical computation;
for applications of this method see \cite{Baker:2003ds,Campanelli:2004zw}.

Progress in the stability properties of 3+1 codes
accelerated considerably in the early 2000s
as a wider range of formulations of the Einstein equations and gauge
conditions were used in BH simulations. The first applications of the
BSSN formulation focused on GW pulses, including collapse to BHs,
and demonstrated significantly
better stability properties than the ADM system
\cite{Shibata:1995we,Baumgarte:1998te,Alcubierre:1999ex}.
Soon afterwards,
this observation was confirmed for evolutions of boson stars,
neutron stars and BHs \cite{Alcubierre:2000xu}.
Using the BSSN formulation combined with a ``$K$-freezing'' slicing
and Gamma-freezing shift (see their Sec.~III for details)
and a ``simple excision'' of a cubic region on whose boundary
time derivatives are copied from neighbouring grid points,
Alcubierre and Br{\"u}gmann \cite{Alcubierre:2000yz} were able to evolve
a Schwarzschild BH in ingoing Eddington Finkelstein
coordinates over many thousands of $M$ with no signs
of instability. These simulations were obtained in 3+1 dimensions
with octant symmetry and the stability properties could be generalized
to 3+1 grids with no symmetry by using the constraint
$\mathcal{G}^i$ of Eq.~(\ref{eq:auxconstraintsBSSN}) in the
evolution equation (\ref{eq:BSSNdtGamma}) \cite{Yo:2002bm,Alcubierre:2002kk}.
Evolutions of distorted BHs with the BSSN system
were pushed to a few $100~M$, about twice
as long as axisymmetric ADM simulations \cite{Alcubierre:2001vm}.
By evolving Brill-Lindquist data with BSSN, 1+log slicing and variants
of the Gamma-freezing shift, combined with the fixed puncture technique
(cf.~Sec.~\ref{sec:BCs}), Alcubierre {\em et al.} \cite{Alcubierre:2002kk}
extracted GWs from BH collisions in good agreement with earlier
axisymmetric ADM simulations but over much extended lifetimes
of $\sim 1\,000~M$. BSSN simulations of head-on collisions
were extended to include mesh refinement in
\cite{Sperhake:2005uf,Fiske:2005fx} using the
{\sc Carpet} \cite{Carpetweb,Schnetter:2003rb} and Paramesh \cite{MacNeice2000}
packages, respectively. The first evolution of a quasi-circular BH binary
extending over more
than one orbital time scale was performed by Br{\"u}gmann {\em et al.}
\cite{Bruegmann:2003aw} in 2003 and explored in more detail
in \cite{Alcubierre:2004hr}. Around the same time BSSN simulations
of inspiraling and merging neutron-star binaries were obtained
by various groups \cite{Marronetti:2003hx,Miller:2003vc,Shibata:2003ga}.
While neutron star spacetimes contain complex matter sources, the
spacetime curvature is significantly weaker than for BH binaries
and there are no singularities (other than nearly stationary
post-merger BHs). Possibly, therein lie the reasons why neutron star
inspiral and merger simulations were achieved before their binary BH
counterparts.
Early 3+1 applications of the GHG formulation focused on the collapse
of scalar fields and the approach to the formation of singularities
using unigrids as well as mesh refinement
\cite{Garfinkle:2001ni,Pretorius:2004jg}.

In 2005, the jigsaw was finally assembled. The first simulations
of BH binaries through inspiral, merger and ringdown were obtained
by Pretorius \cite{Pretorius:2005gq} and a few months later by
the Brownsville/Rochester and NASA Goddard groups
\cite{Baker:2005vv,Campanelli:2005dd}. The new ingredients which
finally pushed the BH simulations ``over the cliff'' can probably
be summarized as follows.
The GHG formulation was adjusted by the addition of
constraint damping terms suggested by Gundlach {\em et al.}
\cite{Gundlach:2005eh}; cf.~Eq.~(\ref{eq:GHG}). Furthermore,
Pretorius managed to specify conditions for the gauge functions
$H^{\alpha}$ that avoided instabilities which had troubled
earlier simulations using the GHG system; cf.~Eq.~(\ref{eq:GHGgauge}).
His compactification of the spacetime and the damping of
signals near spatial infinity through numerical viscosity
prevented outer boundary effects to affect the BH region in
a significant manner. The main new feature of the moving puncture
simulations is encapsulated in the word ``moving''. Up to that point,
puncture simulations had kept the location of the singularity fixed
and factored out the irregular part of the metric analytically.
By using gauge conditions as given in
Eqs.~(\ref{eq:dt_alpha})-(\ref{eq:dt_B}), the Brownsville/Rochester
and NASA Goddard groups obtained a description where the punctures
could freely move across the grid. They furthermore managed to apply
standard finite differencing across the puncture which effectively
provided an exceptionally robust type of singularity excision;
cf.~the discussion in Sec.~\ref{sec:BCs}.

These breakthroughs in simulating BH binaries
triggered a veritable phase transition in the field
of NR as the community gained unprecedented insight into the
dynamics of BH binaries. This is the subject of our next section.





\section{The morphology of black-hole binary inspiral and scattering}
\label{sec:Morphology}

In spite of its dissipative character, the two-body problem in GR
leads to classes of orbits similar, though not identical,
to their Newtonian counterparts discussed at the end of
Sec.~\ref{sec:Newtonian}. The simplest configuration, and focus
of the breakthroughs as well as early follow-up studies,
is the quasi-circular inspiral of two non-spinning BHs of equal
mass. The quasi-circular case is also of particular relevance
for the modeling of GW sources for ground based detectors because
GW emission has long since been known to efficiently decrease the
orbital eccentricity even at large binary separation
\cite{Peters:1964zz}, so that
BH binaries are expected to be circularized to high precision
by the time they reach the frequency window of the detectors.
\begin{figure}[t]
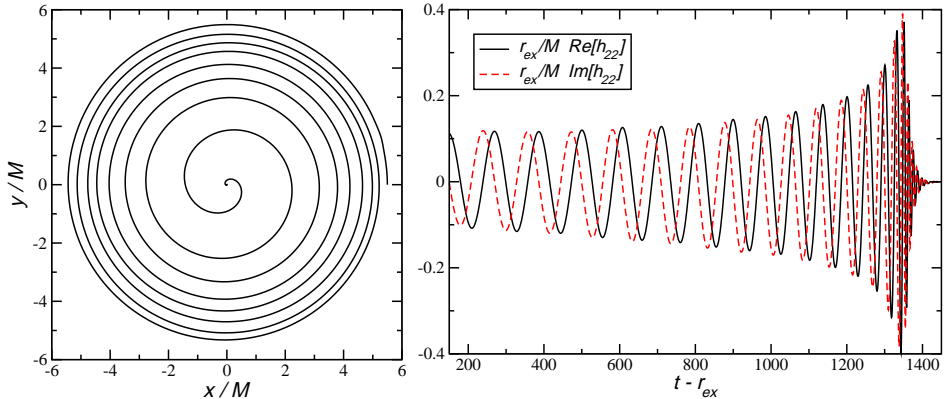

  \centering
  \includegraphics[height=150pt,clip=true,angle=0]{fig5a.eps}
  \includegraphics[height=150pt,clip=true,angle=0]{fig5b.eps}
  \caption{Trajectory (left) and the GW quadrupole (right panel).
           The trajectory of one BH is shown only; the other BH's
           location is obtained through reflection across the origin.
           The waveform shows the real (black, solid) and imaginary
           (red, dashed) part of the strain $h_{22}$ obtained from
           integrating $\psi_{22}$ twice in time.
          }
  \label{fig:q1}
\end{figure}
The dynamics of the late stages of the quasi-circular inspiral and merger
of an equal-mass, non-spinning BH binary are graphically illustrated in
Fig.~\ref{fig:q1}. The left panel shows the trajectory of one BH with that
of the other obtained through reflection at the origin. The binary
separation decreases through the emission of GWs which is dominated
by the quadrupole shown in the form of the GW strain $h$ in the right
panel. The strain is a complex function containing the two GW
polarization modes and related to the Newman Penrose scalar
$\Psi_4$ from which it is calculated by two integrations in time,
and it is the quantity of choice used in GW data analysis; see
e.g.~\cite{Flanagan:1997sx,Flanagan:1997kp}. As the binary
inspirals, the wave signal increases in amplitude and frequency
which reach a maximum around the merger stage followed by the
exponentially damped {\em ringdown} signal; for details
see e.g.~\cite{Buonanno2006,Berti:2007fi}. Eventually, the spacetime
settles down into a stationary Kerr configuration and GW emission ceases.
Realistic astrophysical BH binaries start at a much larger
separation than shown in Fig.~\ref{fig:q1} corresponding to a much longer
inspiral phase of many thousands of cycles \cite{Hinder:2013oqa}.
The breakthrough simulations studied short inspirals of about one
orbit and found that about $3~\%$ of the total mass is radiated in GWs
leaving behind a spinning BH with dimensionless spin parameter of $\sim 0.7$.
These results have been confirmed with good precision by many follow-up
studies, but the earlier inspiral phase not covered in the rather short
simulations provides additional energy release in GWs. The most accurate
simulation to date is that of Scheel {\em et al.} \cite{Scheel2008}
and gives $E_{\rm GW}/M = 0.04838 \pm 0.00002$ for a 16 orbit inspiral
and a dimensionless spin $0.68646\pm0.00004$ for the resulting Kerr BH.

For the reasons mentioned above, eccentric binaries have received less
attention in the NR studies so far. In fact, a good deal of effort
has been spent on measuring the eccentricity in binaries
and reducing the residual eccentricity in the initial data
as much as possible in order to generate high-precision GW templates
for quasi-circular BH systems
\cite{Husa:2007rh,Walther:2009ng,Mroue:2010re,Tichy:2011qa}.
Still, binaries with sizable eccentricity have been studied in their
own right \cite{Hinder:2007qu,Sperhake:2007gu} and also
compared with PN predictions \cite{Hinder:2008kv}. Due to
the GW emission, the eccentricity $\epsilon$ is a time dependent function
in GR; cf.~Eq.~(5.13) in \cite{Peters:1964zz}. For high
values of $\epsilon$, the BHs do not complete a single orbit but
rather plunge into each other. For mild eccentricities, the periastron
advance obtained numerically shows some deviations from
post-Newtonian predictions \cite{Mroue:2010re} but excellent agreement
with perturbative calculations \cite{Tiec:2013twa}.
An interesting feature has been found for eccentric
binaries when the initial momenta of the BHs are fine-tuned, a
{\em threshold of immediate merger} where the binary spends some time
in near-circular orbits before eventually merging or separating
\cite{Pretorius:2007jn}; see also \cite{Gold:2009hr,Gold:2012tk}.
In this regime, also identified in geodesic calculations
\cite{Cutler:1994pb,Glampedakis:2002ya}, the number of {\em zoom-whirl
orbits} exhibits a logarithmic dependence on the deviation of
the impact parameter $b=L/P$ ($L$ is the initial orbital angular momentum
of the binary and $P$ the momentum of either BH in the centre-of-mass frame)
from the threshold of immediate merger
$b^*$ \cite{Pretorius:2007jn,Sperhake:2009jz}. For a fixed value of the
momentum parameter $P$, numerical studies have revealed three possible
regimes separated by two special values of the impact parameter,
the threshold of immediate merger $b^*$ and the {\em scattering threshold}
$b_{\rm scat}$: (i) Prompt mergers resulting from a plunge or
inspiral for $b<b^*$, (ii) non-prompt mergers where the BHs experience
a close encounter, then separate but loose enough energy in GWs to eventually
form a bound system and merge for $b^*< b <b_{\rm scat}$, and (iii)
scattering configurations for $b>v_{\rm scat}$ where the BHs separate
to infinity \cite{Shibata:2008rq,Sperhake:2009jz}.
Here, the latter two regimes can only
be obtained for sufficiently large $P$. We note the similarity between
configurations with $b_{\rm scat}$ and parabolic orbits in Newtonian gravity
which forms the boundary between bound and unbound configurations.
Furthermore, BH collisions at velocities close to the speed of light
can generate enormous amounts of gravitational radiation of up
to $\sim 50~\%$ of the total centre-of-mass energy
\cite{Sperhake:2012me} making them ideal scenarios to probe GR in its
most violent regime.

Unequal mass-ratios $q=m_2/m_1$ (here defined such that $q\le 1$) and/or
non-zero BH spins naturally affect the dynamics of the BHs interaction,
but as yet there are no indications that the above picture of the
different types of orbits is changed in a qualitative manner. For
many practical applications, however, spins and unequal masses have
vital implications. Unequal mass ratios lead to asymmetric emission
of GWs which imparts a recoil or {\em kick} on the post-merger remnant
BH of up to $175~{\rm km/s}$ \cite{Gonzalez:2006md,Baker:2006vn,Herrmann:2007zz}
and specific spin-orientations can lead to so-called {\em superkicks}
of several thousand km/s
\cite{Gonzalez:2007hi,Campanelli:2007ew,Campanelli:2007cga,Lousto:2011kp}.
Spins aligned with the orbital angular momentum cause a so-called
{\em hang-up} effect extending the late inspiral stage and leading
to an increase in GW emission by about a factor of two compared with
the non-spinning case \cite{Campanelli:2006uy,Lovelace:2011nu}.
Signatures of spins and mass ratio furthermore leave specific imprints
such as amplitude modulation or relatively enhanced higher-order
multipoles, which are important effects in the analysis of observational
data of GW detectors and, hence, require extensive modeling combining
NR methods with post-Newtonian and/or perturbative techniques;
see \cite{Ajith:2012tt,Hannam:2013oca,Hinder:2013oqa,Taracchini:2013rva,Damour:2014sva}
and references therein.

It is well beyond the scope of this article to discuss all these
features and the wide range of applications of NR simulations of BH binaries
in detail. Instead, we will refer the interested reader
in the next section to various reviews
that have appeared over the past decade and provide extended
overviews of the many, exciting and sometimes unexpected
developments that have been triggered
by the 2005 breakthroughs in the modeling of BH binaries in GR.






\section{Conclusions}
\label{sec:conclusions}

The 2005 breakthroughs in the modeling of binary BHs mark a phase
transition in the field of NR. Even though some very specific BH systems
(e.g.~head-on collisions) had been modeled successfully before, these
were few isolated and idealized examples. In 2005, virtually
the entire space of BH binary systems was opened up for accurate,
quantitative modeling. The years following 2005 have sometimes been
referred to as the ``goldrush years of numerical relativity'' as the
community suddenly had available the tools to study a wealth
of phenomena previously subject to speculations rather than
precision studies. BH kicks, for example, had been known for decades
to result from GW emission with potentially dramatic astrophysical
consequences but the magnitude of this effect remained largely shrouded
in mystery. The superkicks of thousands of km/s have been one of the
most remarkable and unexpected results of post-2005 NR, a fact still
in the process of digestion in the interpretation of astrophysical
observations \cite{Komossa:2012cy}.

A lot of motivation for the long-term effort of NR came from the
modeling of GW sources in support of the ongoing search for direct
detection with detectors such as LIGO, VIRGO, GEO600 or KAGRA. Here the impact
of the NR breakthroughs is both short and long-term. A wealth of numerical
studies has been devoted to exploring the parameter space of BH binaries
in the attempt to construct waveform catalogues for use in GW data
analysis. The parameter space, however, has at least seven dimensions
(mass ratio and three spin parameters for each BH) making a dense coverage
in every dimension prohibitively costly from a computational point of view.
Instead, numerical as well as analytic studies
have looked for systematic dependencies of
the GW signals on some of the parameters and thus effectively reduce
the parameter space to be explored numerically;
see e.g.~\cite{Apostolatos:1994mx,Brown:2012gs,Schmidt:2014iyl}.
At the same time, the codes have matured considerably in accuracy
and efficiency and are now capable of generating large numbers of
waveforms \cite{Hinder:2013oqa,Mroue:2013xna,Healy:2014eua}.
And yet, a comprehensive modeling of the BH parameter space
certainly cannot be done exclusively with NR because the inspiral
signal of a BH binary in the sensitivity band of ground (or space based)
detectors typically contains many thousands of orbits rather than
the $\mathcal{O}(10)$ orbits covered in numerical simulations. Waveform
templates are therefore constructed by stitching together post-Newtonian
with numerical waveforms as for example in \cite{Ajith:2012tt}
or using {\em effective one body} models \cite{Buonanno:1998gg,Buonanno:2000ef}
and calibrating free parameters by comparison with NR predictions;
see e.g.~\cite{Taracchini:2013rva,Damour:2012ky}. Furthermore, an important
class of sources for space based laser interferometers of LISA type
are so-called extreme-mass-ratio inspirals with mass ratios down
to $\mathcal{O}(10^{-6})$ which the current NR codes
cannot handle; the smallest mass ratios achieved to date
is $q=1/100$ for a small number of orbits \cite{Lousto:2010tb}.
These as well as intermediate mass ratio inspirals with
$q=\mathcal{O}(10^{-3})$ require modeling through perturbative
techniques; see \cite{Poisson:2011nh,Barausse:2011dq} and references therein.

In recent years, BH binary codes have been extended to include
various forms of matter with particular focus on the identification
of electromagnetic counterparts to GW signals. As an example,
we note the variation of the Blandford-Znajek effect in the inspiral
of two BHs in the presence of a circumbinary disk whose magnetic
field is capable of extracting rotational energy from the orbital
motion of the binary giving rise to single or dual jets
\cite{Palenzuela:2009hx,Palenzuela:2010nf}. Other studies address
accretion of matter onto BHs,
the impact of a recoiling BH on the surrounding disk material,
periodic oscillations in the matter due to the binary motion
or the tidal disruption of white dwarfs in the field of a BH
\cite{Bode:2011tq,Alic:2012df,Farris:2012ux,Haas:2012bk};
for an overview see also \cite{Andersson:2013mrx}.
Non-vacuum spacetimes also include some classes of alternative theories
of gravity. NR simulations of BH binaries
have so far focused on scalar-tensor
theories of gravity which are conveniently implemented in existing
GR codes by merely adding a minimally coupled scalar field in the
so-called {\em Einstein frame} \cite{Healy:2011ef,Berti:2013gfa}.
Extension of these studies to a wider class of alternative theories
of gravity may be a highly non-trivial task as the hyperbolicity
properties of the underlying evolution equations and, thus, their
suitability for numerical treatment remain unclear for most theories;
for such an investigation into Dynamical Chern-Simons theory see
\cite{Delsate:2014hba}.

One of the most remarkable developments of NR in the last $\sim 10$
years is the wide range of applications in areas of physics
far outside the more traditional regime of GW and astrophysics.
High-energy collisions of BHs may provide valuable inside into the
cross section of proton-proton collisions performed at the Large Hadron
Collider to probe the possibility of BH formation as conjectured
\cite{Argyres:1998qn,Banks:1999gd,Dimopoulos:2001hw}
in the so-called TeV gravity scenarios
\cite{Antoniadis:1998ig,ArkaniHamed:1998rs,Randall:1999ee,Randall:1999vf}.
The gauge-gravity duality, also often referred to as the
Anti de-Sitter/Conformal Field Theory (AdS/CFT) correspondence
\cite{Maldacena:1997re,Witten:1998qj,Aharony:1999ti} relates BH spacetimes
in asymptotically AdS spacetimes to physical systems in the strongly
coupled regime of certain gauge theories. To name but a few
examples, this duality has been
used to model the thermalization of balls of deconfined
plasma in heavy-ion collisions (see
\cite{Chesler:2008hg,Chesler:2010bi,Heller:2011ju,Chesler:2013lia,Wu:2011yd}
and references therein), the optical conductivity of so-called
strange metals \cite{Horowitz:2012gs,Horowitz:2012ky} or condensed
matter systems (see \cite{Hartnoll:2009sz} for an overview).
Numerical studies of BHs and their formation still teaches us unexpected
lessons about the fundamental properties of general relativity as for example
in Choptuik's critical collapse study \cite{Choptuik:1992jv}
or the surprising instability observed in the extension of Choptuik's
work to asymptotically AdS spacetimes by Bizo{\'n} and Rostworowski
\cite{Bizon:2011gg}. Tests of the cosmic censorship conjecture
are now possible over a wide range of physical scenarios including
high-energy collisions of BHs \cite{Sperhake:2008ga},
BH strings in higher dimensions \cite{Lehner:2010pn} or
BH collisions in asymptotically de Sitter spacetimes
\cite{Zilhao:2012bb}. The numerical simulation of lattices consisting
of multiple BHs are being used to model large scale structures
in cosmological spacetimes \cite{Bentivegna:2012ei,Yoo:2013yea}
and NR tools are developed for the modeling of the early
universe \cite{Garrison:2012ex}.

This short (and incomplete) summary demonstrates that the breakthroughs
of 2005 have opened the door onto a vast garden of opportunities
probably well beyond the wildest expectations the NR community has
harbored during the 40 year path towards the holy grail.
We conclude here
with a list of suggestions for further reading on the various topics
whose surfaces have been scratched on the preceding pages.

Extended books on the methodology of NR have been written
by
Alcubierre \cite{Alcubierre2008},
Bona {\em et al.}~\cite{Bona2009}
as well as
Baumgarte and Shapiro \cite{Baumgarte2010}; we also note
Gourgoulhon's review of the ``3+1'' formulation of GR
\cite{Gourgoulhon:2007ue} and the comprehensive review
by Centrella {\em et al.} \cite{Centrella:2010mx}.
An earlier review also covering the techniques for simulating matter
is given by Lehner \cite{Lehner:2001wq}.
The field of GW physics and the use of BH binary simulations
therein is discussed in various articles
\cite{Hannam:2009rd,Hinder:2010vn,Ohme:2011rm,Pfeiffer:2012pc}.
Comparisons of GW signals obtained numerically with those from
various semi-analytic calculations are reviewed in
\cite{Tiec:2014lba}.
The reader will find overviews of several applications including
astrophysics and high-energy physics in \cite{Pretorius:2007nq,
Sperhake:2011xk}. NR applications outside the more traditional
areas of astrophysics and GW physics are the focus of
\cite{Cardoso:2012qm} and in particular the review
\cite{Cardoso:2014uka}. A description of
more technical details
of BH simulations in asymptotically AdS spacetimes is
given in \cite{Chesler:2013lia}; see also \cite{Bantilan:2012vu}.

If the past $\sim 10$ years of NR have shown anything it is an
astounding potential to surprise us with unexpected results
and new areas of applications. Aside from the above mentioned
articles in the present literature, the reader will undoubtedly
be richly rewarded by following
online and print journals to remain up to date on the avalanche
of results and applications of BH simulations triggered by
the milestone of the 2005 breakthroughs.














\ack
The author thanks
Emanuele Berti,
Bernd Br{\"u}gmann,
Vitor Cardoso,
Pau Figueras,
Jose Gonz{\'a}lez,
Leonardo Gualtieri,
Mark Hannam,
Carlos Herdeiro,
David Hilditch,
Sascha Husa,
Bernard Kelly,
Pablo Laguna,
Luis Lehner,
Christian Ott,
Frans Pretorius,
Harvey Reall,
Christian Reisswig,
Erik Schnetter,
Deirdre Shoemaker,
Ken Smith,
Carlos Sopuerta,
Helvi Witek,
and
Miguel Zilh{\~a}o
for many fruitful discussions. This work was supported by
the FP7-PEOPLE-2011-CIG CBHEO Grant No.~293412,
the FP7-PEOPLE-2011-IRSES NRHEP Grant No. 295189,
the STFC Grant Nos.~ST/I002006/1 and ST/L000636/1,
the Trestles system of the San Diego Supercomputing Centre (SDSC),
Stampede of the Texas Advanced Computing Center (TACC)
and Kraken of the National Centre for Supercomputing Applications (NCSA) through
XSEDE Grant No.~PHY-090003 by the National Science Foundation,
the COSMOS Shared Memory system at DAMTP, University of Cambridge, operated on
behalf of the DiRAC HPC Facility and funded by BIS National E-infrastructure
capital Grant No.~ST/J005673/1 and STFC Grant Nos.~ST/H008586/1
and ST/K00333X/1,
the Centro de Supercomputaci{\'o}n de Galicia (CESGA)
under Grant No.~ICTS-2013-249,
and
the European Union's FP7 ERC Starting Grant DyBHo-256667.


\section*{References}
\bibliographystyle{iopart-num}
\providecommand{\newblock}{}


\end{document}